\documentclass[11pt]{article}
\usepackage{graphicx}
\graphicspath{{./figures/}}
\usepackage{appendix}
\usepackage{latexsym,amsmath,amsfonts,amssymb,booktabs}
\usepackage[font=small]{caption}
\usepackage{slashed,upgreek,amscd,cancel,tensor,color}
\usepackage{adjustbox}
\usepackage[numbers,compress,square]{natbib}
\usepackage{epsfig,latexsym}
\usepackage{amsmath}
\usepackage[pdfencoding=auto]{hyperref} 
\usepackage{url}
\numberwithin{equation}{section}% numera le equazioni seconde le sezioni , e.g. 1.15 invece che consecutivamente; anche le appendici, eq. (A.1) etc. Richiede amsmath
\usepackage{doi}
\usepackage{subcaption}
\definecolor{MyBlue}{rgb}{0.15,0.15,0.70}

\hypersetup{
colorlinks=true,
citecolor=MyBlue,
linkcolor=MyBlue,
urlcolor=MyBlue
}

\input epsf
\usepackage{amssymb}
\usepackage{amsmath}
\usepackage{amsfonts}
\usepackage{upgreek}
\usepackage{latexsym}
\usepackage{stfloats}
\usepackage{afterpage}

\newcommand{\be}{\begin{equation}}
\newcommand{\ee}{\end{equation}}
\newcommand{\beq}{\begin{equation}}
\newcommand{\eeq}{\end{equation}}
\newcommand{\bea}{\begin{eqnarray}}
\newcommand{\eea}{\end{eqnarray}}

\usepackage[stable]{footmisc}

\def\dkmu2{\delta K_{\mu \nu}\delta K^{\mu \nu}}
\def\pmu2{  \phi_{\mu \nu}\phi^{\mu \nu}}

\renewcommand\[{\left[}
\renewcommand\]{\right]}

\newcommand\ees{\end{eqnarray}}
\newcommand\bees{\begin{eqnarray}}

\newcommand{\unitsk}{ \, h \, \text{Mpc}^{-1}}
\newcommand{\unitsr}{ \, h^{-1} \text{Mpc}}
\newcommand{\kvec}{\vec{k}}
\newcommand{\qvec}{\vec{q}}

\newcommand{\eqn}[1]{eq.~(\ref{#1})}

 \newcommand{\momspmeas}[1]{\frac{d^3 #1}{(2 \pi)^3}}
 \newcommand{\half}{\frac{1}{2}}

\newcommand{\xvec}{\vec{x}}
\newcommand{\knl}{k_{\rm NL}}

\newcommand{\secref}[1]{Sec.~\ref{#1}}
\newcommand{\footnoteref}[1]{Footnote~\ref{#1}}
\newcommand{\appref}[1]{App.~\ref{#1}}
\newcommand{\figref}[1]{Fig.~\ref{#1}}

\newcommand{\im}{\text{Im}}
\newcommand{\mo}{m_1}
\newcommand{\mt}{m_2}

\newcommand{\rvec}{\vec{r}}
\newcommand{\yvec}{\vec{y}}
\newcommand{\nuo}{\nu_1}
\newcommand{\nut}{\nu_2}

\newcommand{\diffsq}{| \vec{k} - \vec{q}|^2}

\newcommand{\citees}[1]{\cite{#1}}
\newcommand{\pvec}{\vec{p}}
\newcommand{\es}{\epsilon_{s}}
\newcommand{\ed}{\epsilon_{\delta}}
\newcommand{\esp}{\epsilon_{s}'}
\newcommand{\lbao}{\ell_{\rm BAO}}
\newcommand{\esl}{\epsilon_{s<}}

\begin{document}
\vspace{0.5cm}

\vspace{0.5cm}

\begin{center}
\Large{\textbf{An analytic implementation \\ of the IR-resummation for the BAO peak}} \\[1cm]

\large{Matthew Lewandowski$^1$ and Leonardo Senatore$^{2,3}$}
\\[0.5cm]

\small{
\textit{$^1$ Institut de physique th\' eorique, Universit\'e  Paris Saclay, \\ 
CEA, CNRS, 91191 Gif-sur-Yvette, France  }}

\vspace{.2cm}

\small{
\textit{$^{2}$ Stanford Institute for Theoretical Physics,\\ Stanford University, Stanford, CA 94306 }}

\vspace{.2cm}
\small{
\textit{$^{3}$ Kavli Institute for Particle Astrophysics and Cosmology, \\
Physics Department and SLAC, Menlo Park, CA 94025 }}

\vspace{.2cm}

\vspace{0.5cm}
\today

\end{center}

\vspace{2cm}

\begin{abstract}
We develop an analytic method for implementing the IR-resummation of \cite{Senatore:2014via}, which allows one to correctly and consistently describe the imprint of baryon acoustic oscillations (BAO) on statistical observables in large-scale structure.   We show that the final IR-resummed correlation function can be computed analytically without relying on numerical integration, thus allowing for an efficient and accurate use of these predictions on real data in cosmological parameter fitting.  In this work we focus on the one-loop correlation function and the BAO peak.  We show that, compared with the standard numerical integration method of IR-resummation, the new method is accurate to better than $0.2\%$, and is quite easily improvable.  We also give an approximate resummation scheme which is based on using the linear displacements of a fixed fiducial cosmology, which when combined with the method described above, is about six times faster than the standard numerical integration.  Finally, we show that this analytic method is generalizable to higher loop computations.
\end{abstract}

\newpage

%\newpage
\tableofcontents

\vspace{.5cm}
\newpage

%%%%%%%%%%%%%%%%%%%
%
%

%%%%%%%%%%%%%%%%%%%

\section{Introduction}

Upcoming large-scale structure (LSS) surveys have the potential to be the next leading sources of cosmological information.  This is because the amount of information contained in these three-dimensional surveys scales roughly like $k_{\rm max}^3$, where $k_{\rm max}$ is the maximum wavenumber that is under theoretical control.  This is to be contrasted with the amount of information in two-dimensional measurements, like the cosmic microwave background (CMB), which scales like $k_{\rm max}^2$.  Unfortunately, for the CMB, there is a physical upper limit for $k_{\rm max}$ due to Silk damping that makes it difficult to improve the reach by a large amount beyond already existing analyses.  Thus, we are compelled to have a strong theoretical understanding of gravitational clustering at the highest wavenumber (or smallest length scales) possible.  In order to do this, the Effective Field Theory of Large-Scale Structure (EFTofLSS) was introduced \cite{Baumann:2010tm,Carrasco:2012cv} and subsequently developed \cite{Carrasco:2013mua,Carrasco:2013sva,Porto:2013qua,Pajer:2013jj,Carroll:2013oxa,Mercolli:2013bsa,Angulo:2014tfa,Senatore:2014via,Baldauf:2014qfa,Senatore:2014eva,Senatore:2014vja,Lewandowski:2014rca,Mirbabayi:2014zca,Foreman:2015uva,Angulo:2015eqa,McQuinn:2015tva,Assassi:2015jqa,Baldauf:2015tla,Baldauf:2015xfa,Foreman:2015lca,Baldauf:2015aha,Baldauf:2015zga,Bertolini:2015fya,Bertolini:2016bmt,Assassi:2015fma,Lewandowski:2015ziq,Cataneo:2016suz,Bertolini:2016hxg,Fujita:2016dne,Perko:2016puo,Lewandowski:2016yce,Senatore:2017hyk,delaBella:2017qjy,Nadler:2017qto,Foreman:2018gnv,deBelsunce:2018xtd} to consistently describe clustering observables in the mildly non-linear regime.  This list includes studies of the dark-matter power spectrum, bispectrum, and trispectrum, lensing, redshift space distortions, biased tracers, the baryon acoustic oscillation (BAO) peak, the effects of baryons and massive neutrinos, dark energy, and non-gaussianity, thus putting the EFTofLSS in a position to be able to be applied directly to the analysis of survey data.

In general, cosmological perturbation theory is a loop expansion, using Green's functions, which accounts for the effects of the non-linear evolution of initial conditions (see \cite{Bernardeau:2001qr} for a review of standard perturbation theory (SPT)).  The loop corrections involve integrals over intermediate momenta, and in particular, over short-scale (i.e. ultraviolet or UV) momenta for which the equations of the theory are not specifically known.  To account for the unknown physics at short scales, the EFTofLSS introduces counterterms which allow for the effects of UV modes on large-scale observables to be consistently incorporated in the perturbative expansion.  Once this is done, the result is a controlled expansion in $k / \knl$, where $\knl$ is the strong coupling scale of the EFT (or equivalently the non-linear scale of structure formation).  For modes with $k \lesssim \knl$, observables can be predicted to higher and higher accuracy (up to non-perturbative effects) by including more and more loops.  

\begin{figure}[htb!]
\includegraphics[width=15cm]{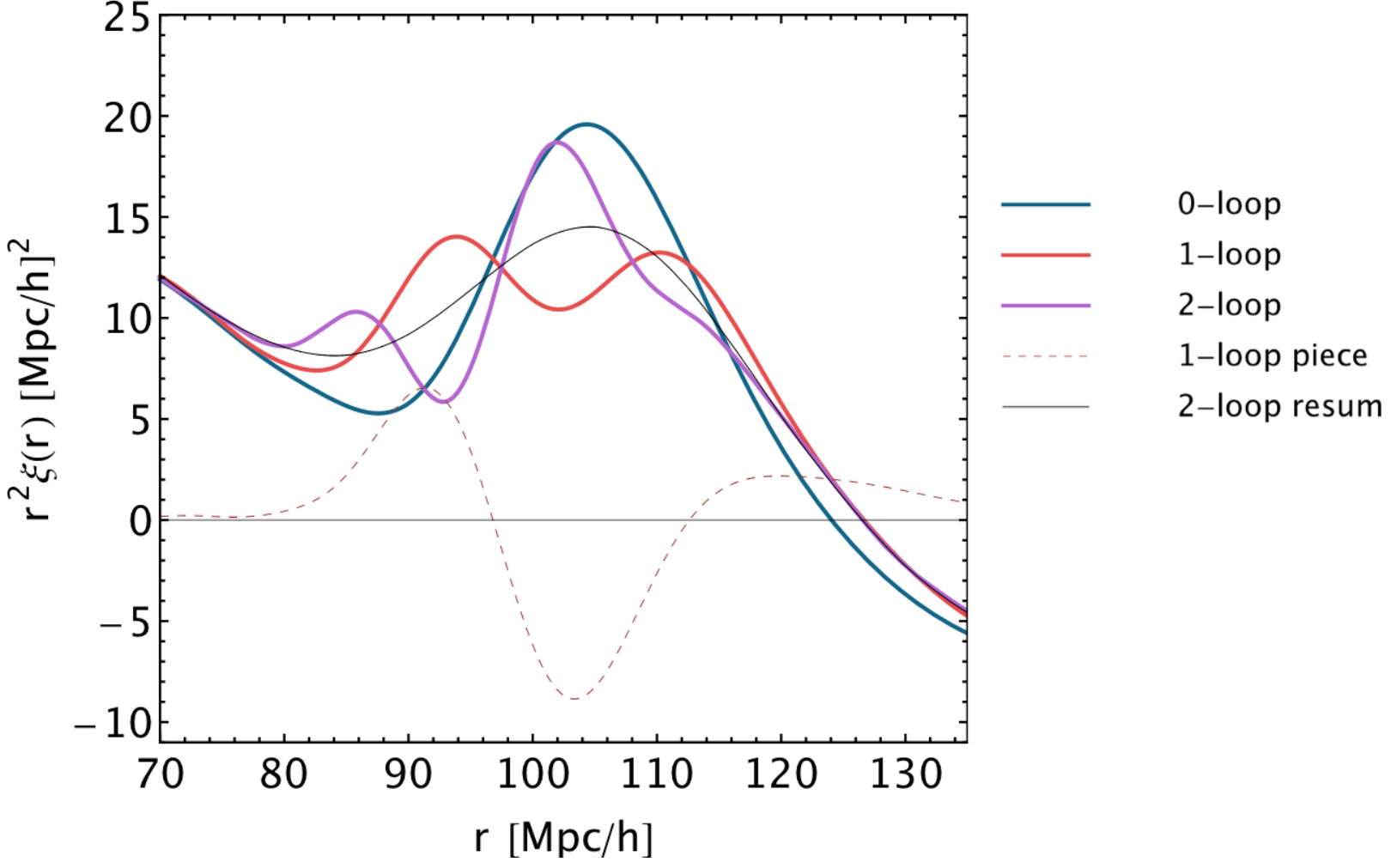} 

\caption{The BAO peak in Eulerian perturbation theory at tree level (dark blue), one loop (red), and two loops (purple), along with the individual one-loop contribution (dashed red).  We see that the convergence is quite slow since the difference between the tree-level and one-loop curves is about the same as the difference between the one-loop and two-loop curves.  In this plot, we also show the two-loop resummed correlation function (thin black) as a proxy for the correct answer, see \secref{resultssec} for a more thorough discussion.  } \label{introplot}
\end{figure}

The perturbative expansion can be done in two main frameworks: Eulerian perturbation theory or Lagrangian perturbation theory (see \cite{Bernardeau:2001qr} for reviews of these approaches in SPT).  Most of the developments in the EFTofLSS have been presented in the Eulerian framework, although there have been some developments in the Lagrangian framework \cite{Porto:2013qua,Vlah:2015sea}.  The main difference between the two is that the fundamental quantities in Eulerian perturbation theory are local fields (like the over-density $\delta(\xvec)$ and velocity $v^i ( \xvec)$) in the fluid picture, whereas Lagrangian perturbation theory expands around a fluid element's full trajectory.  Because of this, the Eulerian picture is not as good at describing bulk (i.e. infrared or IR) displacement effects.  While the EFTofLSS in the Eulerian picture correctly deals with the UV and improves the maximum reach in $k$ space, it has been known that it does not converge rapidly when describing the BAO peak (see for example \cite{Crocce:2007dt}), see \figref{introplot}.  This is due to Eulerian perturbation theory's expanding in the long displacements, which for our universe are order one, and so convergence is slow at best. 

More specifically, we can look at the parameters that are relevant to the loop expansion of Eulerian perturbation theory \cite{Senatore:2014via,Baldauf:2015xfa,Tassev:2013rta}
\begin{align} \label{epsilondef}
 \epsilon_{s < } ( k )  \equiv  k^2 \int_0^k \frac{d^3 k'}{( 2 \pi)^3} \frac{ P_{11} ( k ' )}{k'^2}   \ , \hspace{.07in}   \epsilon_{\delta < }  ( k ) \equiv  \int_0^k \frac{d^3 k'}{( 2 \pi)^3} P_{11} ( k ' )  \ , \hspace{.07in} \text{and} \hspace{.07in}  \epsilon_{s >} ( k ) \equiv  k^2 \int_k^\infty \frac{d^3 k'}{( 2 \pi)^3} \frac{ P_{11} ( k ' )}{k'^2}  \ ,
\end{align}
where $\langle \delta^{(1)} ( \kvec) \delta^{(1)} ( \kvec ') \rangle = (2 \pi)^3 \delta_D ( \kvec + \kvec ' ) P_{11} ( k )$ defines the linear power spectrum $P_{11}$ in terms of the linear solution $\delta^{(1)} ( \kvec)$.  
As is evident in \eqn{epsilondef}, $\epsilon_{s <}$ is related to IR displacements, $\epsilon_{ \delta < }$ is related to IR density fluctuations, and $\epsilon_{s >}$ is related to UV displacements.  Eulerian perturbation theory expands equally in all three of these parameters.  However, linear displacements make $\epsilon_{s<}$ order one on the scales of interest, which affects the success of perturbation theory for describing the BAO peak and non-IR-safe quantities.\footnote{For IR-safe quantities, the relevant parameter is actually $\epsilon_{s<}^{\rm BAO}$, which is defined the same as $\epsilon_{s<}$ in \eqn{epsilondef}, but with the integral going from $k_{\rm BAO} \sim 2 \pi / \ell_{\rm BAO}$ (where $\ell_{\rm BAO} \approx 100 \unitsr$ is the position of the BAO peak) to $k$.  Quantitatively, $\epsilon_{s<}^{\rm BAO} \simeq \epsilon_{s<}$. }  When computing equal-time correlation functions, it is known that effects from deep IR modes must cancel  \cite{Jain:1995kx, Scoccimarro:1995if,Peloso:2013zw,Carrasco:2013sva}.  However, for IR modes around the scale of the BAO oscillations, this cancellation does not happen, and because $\epsilon_{s<}^{\rm BAO}$ is large, the perturbative expansion fails to capture the BAO oscillations.  

To deal with this, a controlled, hybrid Eulerian and Lagrangian picture IR-resummation was developed~\cite{Senatore:2014via} (see also \cite{Matsubara:2007wj,Lewandowski:2015ziq,Senatore:2017pbn} for related work).  The IR-resummation is a method to treat fully non-linearly the effects of IR displacements and resum the parameter $\epsilon_{s<}$ (for recent related methods, see \citees{Baldauf:2015xfa,Vlah:2015sea,Vlah:2015zda,Blas:2016sfa,delaBella:2017qjy,Ivanov:2018gjr}, and for earlier reconstruction methods, see for example \cite{Eisenstein:2006nj,Seo:2008yx}).  In summary, the correlation function at all orders in $\epsilon_{s<}$ and expanded to order $N$ in $\epsilon_{\delta < }$ is given by,\footnote{For the purposes of this paper, we do not distinguish between $\epsilon_{\delta<}$ and $\epsilon_{s>}$, as the effect of both of those is computed perturbatively.}
\be \label{generalresumeq1}
\xi(r )\big|_N =  \sum_{j=1}^N \int d^3 q \,  \xi^{\rm E}_j ( q ) R_{N-j} ( \rvec - \qvec , \rvec) \ ,
\ee
where $\xi^{\rm E}_j ( q )$ is the $j$-th order contribution to the Eulerian correlation function, and the kernels $ R_{N-j}$ contain the information about the long wavelength displacements: the form of the kernels $ R_{N-j}$ is derived from the Lagrangian picture \cite{Senatore:2014via, Senatore:2017pbn}.  Many more details are given in the rest of the paper (see \appref{irresumexpsec}), but now we wish to stress one main point about the form of \eqn{generalresumeq1}.  In order to use \eqn{generalresumeq1}, one must first compute the standard Eulerian loops to compute the $j$-th order piece $\xi^{\rm E}_j ( q )$, then one must do another integral (which turns out to be a one-dimensional integral over $d q$) to apply the IR-resummation.  Typically, these integrals have been done numerically, but this can be rather slow, especially for higher loops and in redshift space.  Ultimately, in order to make this method much more useful for data analyses, it is convenient to have a faster method: speeding up the computation of \eqn{generalresumeq1} is the main goal of this paper.

 It is worth noting that other fast and analytic (or at least semi-analytic), methods of resumming the long-wavelength displacements have also been developed, see for example \cite{Baldauf:2015xfa,Vlah:2015zda,Blas:2016sfa}.  These methods rely on breaking the power spectrum into a smooth part and a wiggle part, which drastically simplifies the expressions of \cite{Senatore:2014via} for the full resummation.  While this allows for a fast computation, the procedure of splitting the power spectrum, along with ignoring certain angular terms, introduces a small error in the computation.  These errors, while small, are parametrically different from the parameters in which we expand in the EFT and in general are only recovered thanks to the slow convergence of the non-resummed perturbation theory.  By using the exact formulas, though, one can just dispense with this issue.  See \appref{wnwcxnsec} for further discussion on this issue.

To accomplish our goal, we start with the fast computation of the power spectrum, which uses the FFTLog and dimensional regularization to compute the Eulerian loops \cite{McEwen:2016fjn,Schmittfull:2016jsw,Schmittfull:2016yqx,Hamilton:1999uv,Simonovic:2017mhp,Scoccimarro:1996jy}.  The end result of this procedure is that one decomposes the linear power spectrum $P_{11}$ into a sum of complex power laws, and for each power law, the one-loop integrals can be done analytically.  Using this, we show that, for each of the complex power laws in which $P_{11}$ is decomposed, the integrals in \eqn{generalresumeq1} can be done analytically.  Thus, the resulting expression for the IR-resummation does not contain any numerical integrals, and we are left with a sum over analytical functions weighted by the coefficients of the decomposition of $P_{11}$ with the FFTLog.  The final method presented in this paper, which is a combination of the fixed-displacements approximation discussed in \secref{fixeddispsec} and the exact evaluation discussed in \secref{exatsec}, is about six times faster than the numerical integral technique.  The accuracy of our approach depends on a few different approximations.  When using the fixed-displacements approximation on cosmologies which have linear power spectra different by a few percent, the error in the one-loop resummation is approximately $0.1\%$, and the error in the two-loop resummation is less than $0.05\%$ (indeed the error is diminished at each loop order).  On the other hand, in order to have the analytical form of the resummation integrals, we have to expand a $\Gamma$ function in the integrand, and this introduces a systematic, although easily improvable, error.  We find that including the first two terms in that $\Gamma$ function expansion, the result is within $0.2\%$ of the numerical integral technique.

The layout of this paper is as follows.  In \secref{decompsec} we review the decomposition of the power spectrum using the FFTLog.  In \secref{analyticresumsec}, we present the analytic IR-resummation for the one-loop correlation function.  In \secref{evalsec} we give three methods to efficiently evaluate our new formulas: using the exact formulas, using the saddle-point approximation, and expanding around the displacements of a fixed cosmology.  Finally, in \secref{resultssec} we present and discuss our results, compare to the non-linear data of the Dark Sky simulation \cite{Skillman:2014qca}, and conclude.\footnote{We have provided the Mathematica notebook developed during this project at the \href{http://stanford.edu/~senatore/}{EFTofLSS repository}.}

%%%%%%%%%%%%%%%%%%%%%%
%
%%

\section{Decomposition of the power spectrum} \label{decompsec}

Following \cite{McEwen:2016fjn, Hamilton:1999uv, Simonovic:2017mhp}, we can decompose the linear power spectrum as a Fourier series in $\log k$.   To reproduce the linear power spectrum at $N_{\rm max}$ points $k_n$, equally $\log$-spaced from $k_{\rm min}$ to $k_{\rm max}$, we can write
\be \label{plindef}
P_{11} ( k_n ) = \sum_{m=-N_{\rm max}/2}^{N_{\rm max}/2}  c_m k_n^{-2 \nu_m}
\ee
where $-2 \nu_m \equiv \nu + i \eta_m$, $\nu$ is a fixed real number called the bias, and the expansion coefficients $c_m$ and frequencies $\eta_m$ are given by
\be \label{cmetam}
c_m = \frac{1}{N} \sum_{l = 0 }^{N_{\rm max} -1} P_{11} ( k_l ) k_l^{-\nu} k_{\rm min}^{- i \eta_m } e^{-2 i m l /N} \ , \quad \quad \eta_m = \frac{2 \pi m}{\log ( k_{\rm max} / k_{\rm min})} \ . 
\ee
This decomposition can be done quickly and efficiently using the FFTLog \cite{Hamilton:1999uv, McEwen:2016fjn}.

With the decomposition given in \eqn{plindef}, the authors of \cite{Simonovic:2017mhp} used the fact that the one-loop integrals (see \appref{oneloopformsec}) for the individual power law terms in $ P_{11}$ can be done analytically to write the one-loop power spectrum as a simple matrix multiplication.  The end result is that the one-loop contributions $P_{13}$ and $P_{22}$ can be written as (below $\sigma \in \{13,22\}$),  
\be \label{p1loopdecomp}
P_{\sigma} ( k ) = k^3 \sum_{m_1, m_2} c_{m_1} k^{-2 \nu_1} M_{\sigma} ( \nu_1,\nu_2 )  k^{-2 \nu_2} c_{m_2}
\ee
where we have introduced the shorthand $\nu_i \equiv \nu_{m_i}$, the one-loop matrices $M_\sigma$ are given by 
\be
M_{22} (\nu_1, \nu_2)= \frac{( \frac{3}{2} - \nu_{12} )(  \frac{1}{2} - \nu_{12} ) [ \nu_1 \nu_2 ( 98 \nu_{12}^2 - 14 \nu_{12} + 36 ) - 91 \nu_{12}^2  +3 \nu_{12} + 58    ]   }{196 \nu_1 ( 1 + \nu_1)(\half - \nu_1) \nu_2 (1+\nu_2)(\half - \nu_2)} \textsf{I} ( \nu_1 , \nu_2) \ , 
\ee
\be
M_{13} ( \nuo , \nut ) = \frac{1 + 9 \nuo}{4} \frac{ \tan ( \nuo \pi )}{28 \pi (\nuo + 1) \nuo (\nuo-1) (\nuo -2) (\nuo -3)} \ ,
\ee
with\footnote{Here, the $\Gamma$ function is given by $\Gamma ( s, x ) =\int_x^\infty dt \, t^{s-1} e^{-t}$ and $\Gamma( s ) \equiv \Gamma(s,0)$. \label{gammaftnt}}
\be 
\textsf{I} ( \nu_1 , \nu_2)  = \frac{1}{8 \pi^{3/2}} \frac{\Gamma ( \frac{3}{2} - \nu_1) \Gamma ( \frac{3}{2} - \nu_2) \Gamma ( \nu_{12} - \frac{3}{2} ) }{ \Gamma( \nu_1) \Gamma ( \nu_2) \Gamma ( 3 - \nu_{12}) } \ , 
\ee
and we have used the notation $\nu_{12\dots n} \equiv \nu_1 + \nu_2 + \cdots + \nu_n$.  This expression comes from using dimensional regularization to do the following integral which is present in the one-loop computation \cite{Scoccimarro:1996jy,Pajer:2013jj}  
\be \label{dimreg}
\int \momspmeas{q} \frac{1}{q^{ 2 \nuo} |\kvec - \qvec|^{2 \nut}} \equiv k^{3 - 2\nu_{12}} \textsf{I} ( \nu_1 , \nu_2) \ . 
\ee
Finally, the one-loop counterterm is given by 
\be
P_{13}^{\rm ct} ( k ) = -4 \pi c_s^2 \sum_{m} c_{m} k^{-2(\nu_m -1)} \ .  
\ee
For details concerning convergence of the loop integrals and the use of \eqn{dimreg}, see \appref{oneloopformsec}.  The end result is that, for the simplest application of the above formulas, one should consider $-3 < \nu $, although the adjustments to be made if $\nu$ is outside of this range are straightforward \cite{Simonovic:2017mhp}.

% dimregexpr

%%%%%%%%%%%%%%%%%%
%
%
%
%
%%%%%%%%%%%%%%%%%%

\section{An analytic implementation of the IR-resummation} \label{analyticresumsec}
%%%%%%%%%%%%
%
%
%

In this section, we focus on the one-loop IR-resummation (expressions related to higher loop expressions can be found in \appref{higherloopssec}).  As mentioned in the Introduction, the standard Eulerian perturbation theory is an expansion in IR density fluctuations $\epsilon_{\delta < }$, IR displacements $\epsilon_{s<}$, and UV displacements $\epsilon_{s >}$, which are all scale dependent \cite{Senatore:2014via}.  A drawback of this expansion is that the IR displacements become large on the scales of interest, and the Eulerian expansion is unable to correctly describe the BAO oscillations \cite{Crocce:2007dt}.  On the other hand, Lagrangian perturbation theory, which follows individual fluid elements, is much better at describing these bulk motions (see e.g. \cite{Matsubara:2007wj}).  With this in mind, \cite{Senatore:2014via} developed a hybrid perturbation theory, called the IR-resummation, which treats the linear IR displacements non-perturbatively and then expands the rest as in Eulerian perturbation theory.  This procedure resums most of the IR displacements; the effects from the remaining IR displacements, which are expanded in the loop expansion, are then characterized by a new parameter $\tilde \epsilon_{s <} \ll 1 \lesssim \epsilon_{s <}$, and so are perturbative and captured order by order in the loop expansion.  In this paper, we follow this method to resum the IR modes.  Related methods have been proposed in \cite{Baldauf:2015xfa,Vlah:2015sea,Vlah:2015zda,Blas:2016sfa,delaBella:2017qjy,Ivanov:2018gjr}.

%%%%%%%%%%%%%%%%%%%%%%%

\subsection{Decomposition of the correlation function}

In this paper, we will work mostly with the correlation function $\xi ( r )$, which is defined by 
\be \label{corrfndef}
\xi( r ) = \int_0^\infty d k \frac{k^2}{2 \pi^2} j_0 ( k r ) P(k) \ ,
\ee
where $j_0 ( x ) \equiv \sin(x) / x $ is the spherical Bessel function of index $0$.  It turns out that with the decomposition of the power spectrum given in \secref{decompsec}, we can analytically compute the spherical Bessel transform in \eqn{corrfndef}.  This is not a surprise, however, since this is how the spherical Bessel transform (SBT) based on the FFTlog works \cite{Hamilton:1999uv}.  Using the fact that\footnote{This is a specific case of the more general identity
\be
\int_0^\infty d x \, j_\mu ( x ) x^{2 - 2 \nu} = \sqrt{\pi} \,  2^{1-2\nu} \frac{\Gamma \left[ (3 + \mu -2 \nu) /2 ) \right]}{\Gamma \left[ ( \mu + 2 \nu ) /2 \right]} \ , 
\ee
where $j_\mu ( x )$ is the spherical Bessel function of index $\mu$.}
\be  \label{besseltransf}
 \int_0^\infty d  x \,  j_0 ( x ) \, x^{2 -2 \nu_m} = \Gamma ( 2 - 2 \nu_m) \sin ( \pi \nu_m )  \ , 
 \ee
we find (letting $2 \omega_m = 2 \nu_m -3$ for later convenience),
\be \label{m11tilde}
\xi_{11} ( r ) =   \sum_m  c_m  \tilde M_{11} ( \nu_m) r^{2 \omega_m} \ , \hspace{.2in} \text{with} \hspace{.2in} \tilde M_{11} ( \nu_m)  \equiv \frac{1}{2 \pi^2} \Gamma( 2 - 2 \nu_m) \sin ( \pi \nu_m) \ ,
\ee
and for the one-loop expressions
\begin{align} \label{xioneloopbessel}
\xi_{\sigma} ( r ) & =  \sum_{m_1,m_2} c_{\mo}   r^{2 \omega_1}  \tilde M_{\sigma} ( \nu_1 , \nu_2 ) r^{2 \omega_2} c_{\mt} \ ,  
\end{align}
with 
\be
\tilde M_{\sigma} ( \nu_1 , \nu_2 ) \equiv \frac{1}{2 \pi^2} \Gamma ( 5 - 2 \nu_{12} ) \cos ( \pi \nu_{12} ) M_\sigma ( \nu_1 , \nu_2)  \ .
\ee
Finally, the counterterm contribution is 
\be \label{xicountertermbessel}
\xi^{\rm ct}_{13} ( r ) = - 4 \pi c_s^2 \,  \sum_m c_m \tilde M_{11} ( \nu_m -1 ) r^{2 ( \omega_m -1)} \ . 
\ee

%%%%%%%%%%%
%
%

\subsection{Tree-level resummation}

We start with the tree level resummation; the relevant expressions for the IR-resummation can be found in \appref{irresumexpsec}.  We must compute
\begin{align}
\xi( r ) \big|_0 = \int d^3 q \, R_0 ( \rvec - \qvec , \rvec ) \, \xi_{11}^{\rm E} ( q )  \ ,
\end{align}
where here and elsewhere, the superscript $\rm E$ stands for Eulerian, meaning that the quantity is the standard one computed in Eulerian perturbation theory expanded in both $\epsilon_{\delta <}$ and $\epsilon_{s <}$.
Plugging in the decomposition \eqn{m11tilde} for $\xi_{11}^{\rm E}$, we can write this as
\be \label{treelevelresum1}
\xi( r ) \big|_0 =  \sum_m c_m \tilde M_{11} ( \nu_m) \tilde \xi^{(0)}_{\omega_m} ( r ) 
\ee
where
\begin{align}
\begin{split} \label{tildexizero}
\tilde \xi^{(0)}_{\omega_m} ( r ) &\equiv  \int d^3 q \, q^{ 2 \omega_m} R_0 ( \rvec - \qvec , \rvec)  = \frac{ (2 \pi)^{-3/2}}{\sqrt{| A ( \rvec ) |}} \int d^3 q \, q^{2 \omega_m} \exp \left\{ - \half (\rvec - \qvec)^i A_{ij}^{-1} ( \rvec ) (\rvec - \qvec)^j \right\}    \ .
\end{split}
\end{align}
The reason for the superscript $(0)$ will become evident when we compute the one-loop resummation, see \eqn{xitildelambda} later for example.  To do the integral in \eqn{tildexizero}, we shift integration variable to $\qvec \rightarrow \rvec - \qvec$, then use the rotation invariance of the integral to rotate to integration variables which diagonalize $A^{-1}$ (see \appref{irresumexpsec}) to get
\begin{align} \label{quicksand}
\tilde \xi^{(0)}_{\omega_m} ( r ) = \frac{ (2 \pi)^{-3/2}}{\sqrt{| A ( \rvec ) |}} \int d^3 q \left( q_1^2 + q_2^2 + (r - q_3)^2 \right)^{\omega_m} \exp \left\{  - \half \left( \alpha_0  ( q_1^2 + q_2^2 ) + ( \alpha_0 + \alpha_2  ) q_3^2 \right) \right\} \ ,
\end{align}
where, to avoid clutter, here and below we will often drop the argument $r$ on $\alpha_{0} ( r )$ and $\alpha_2 ( r )$ because these functions are constant for the purpose of the integrals that we are doing: they should all be evaluated at the external point $r$.  
Now, the integrals over $q_1$ and $q_2$ can be done analytically.  This gives 
\begin{align} 
\begin{split}
\tilde \xi^{(0)}_{\omega_m} ( r ) & = \frac{ (2 \pi)^{-3/2}}{\sqrt{| A ( \rvec ) |}}  \int_{-\infty}^\infty d q_3 \, \pi \left( \frac{2}{\alpha_0 } \right)^{1+\omega_m}  (r-q_3)^{2 \omega_m}  \Gamma \left( 1 + \omega_m , \frac{\alpha_0}{2} ( r - q_3)^2 \right) \\
 & \hspace{1.5in} \times \exp \left\{ - \half \left( ( \alpha_0 + \alpha_2  ) q_3^2 - \alpha_0  ( r - q_3)^2 \right) \right\}  \ . \label{intermed}
 \end{split}
\end{align}

\begin{figure}[htb!]
\hspace{1.2in} \includegraphics[width=10cm]{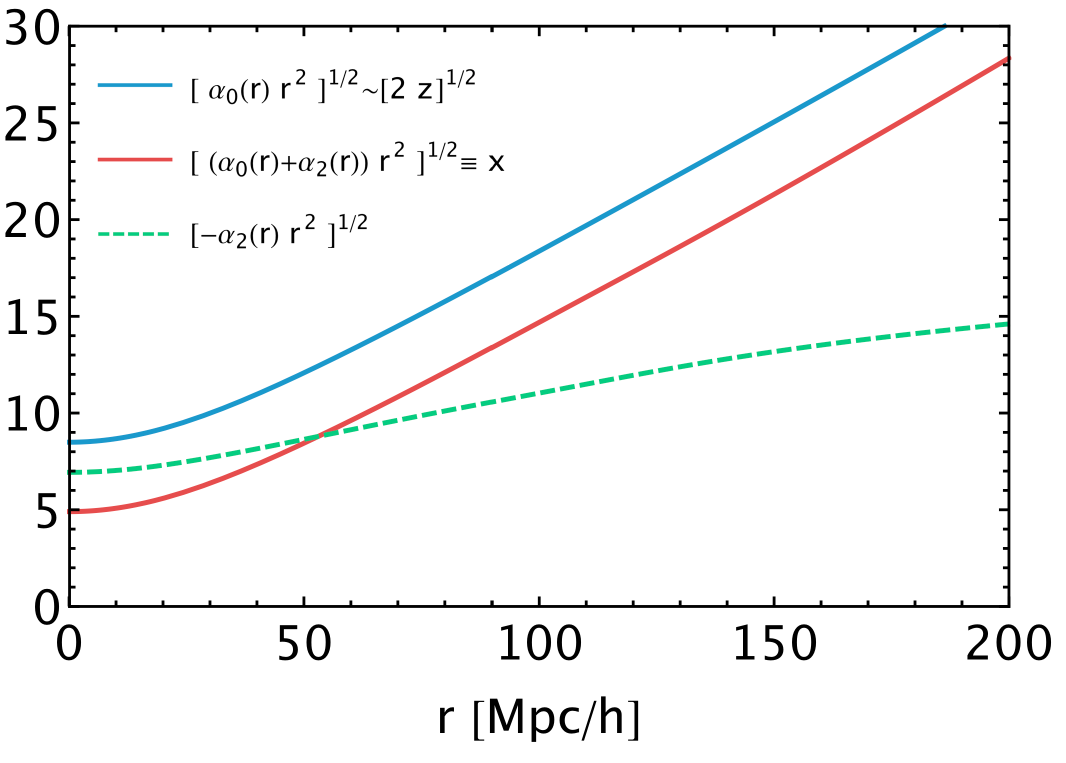} 

\caption{We plot the functions $\alpha_0 (r )$ and $\alpha_2 ( r )$ which are related to the long-wavelength displacements, see \appref{irresumexpsec}, for the cosmology C1 described in \secref{resultssec}.  The forms of the functions plotted here are the ones which are most relevant for the calculations that we do in this paper. } \label{alphaplot}
\end{figure}

In order to do this integral analytically, we can expand the $\Gamma$ function in \eqn{intermed} for the large argument $\alpha_0 ( r - q_3 )^2 /2$.  This is justified because the exponential will force the integral to be dominated by $q_3 \ll r$, and $\alpha_0(r) r^2 \sim 80 $ for $r \approx 1$, and is larger for larger $r$, see \figref{alphaplot}.  As we will see, it is easy to keep as many terms as necessary in this expansion, although as we will discuss later, we find that it is enough for our purposes to keep just the zeroth- and first-order terms.  The expansion of the $\Gamma$ function for $z \rightarrow \infty$ is given by 
\begin{align} 
\begin{split} \label{gammaexp1}
\Gamma( 1 + \omega , z ) & \approx e^{-z} z^\omega \left( 1 + \frac{ \omega}{z} + \frac{-\omega + \omega^2}{z^2} +  \dots \right)  \equiv e^{-z} z^{\omega} \sum_{j=0}^\infty \gamma_j ( \omega ) z^{-j} \ , 
\end{split}
\end{align}
where the coefficients $\gamma_j ( \omega)$ are defined by the expansion in \eqn{gammaexp1}.\footnote{  To be more precise, as we will see, the largest value of $| \omega |$ that we will typically have is $| \omega | \approx 20$, and the smallest value of $z $, which is approximately equal to $ \alpha_0 ( r ) r^2 / 2$, that we will have is 40 (which occurs for $r\rightarrow 0$, although the value near the BAO peak is $ \alpha_0 ( r ) r^2 / 2 \approx  150$, see \figref{alphaplot}).  Thus, the smallest value of $\omega / z$ that we expect to have in \eqn{gammaexp1} is $\omega / z \approx 1/2$, which naively may not seem like a very good expansion parameter.  However, there are two reasons not to be worried.  First, although the series for $\Gamma ( 1 + \omega , z )$ that we use is asymptotic, we find that for the worst value of $\omega/z$ that we use, the maximum precision that the series obtains is better than $10^{-6}$, which is much smaller than any other error discussed in this paper, so one can increase precision by using a higher order expansion.  Second, looking back at \eqn{treelevelresum1}, to get the final answer, we are ultimately summing over many values of $\omega$, and most of them have a smaller ratio of $\omega/z$ anyway.  So, the ultimate precision of this method is a number that is much better than $10^{-6}$, as terms with $\omega / z \sim 1/2$ do not make up most of the overall terms.  As a final note, other series representations of the $\Gamma$ function appear in the literature which may have better properties, but we find \eqn{gammaexp1} more than precise enough for our purposes.  In any case, this $10^{-6}$ precision is such a high level of accuracy that we can worry about this when we have computed the answer to such a high loop order that $\epsilon_{\delta <}$ corrections are comparable to this, or when we have observational data that match this precision (which does not appear to be any time soon). }  
This then allows us to write 
\begin{align}
 \tilde \xi^{(0)}_{\omega_m} ( r )  = \sum_j   \gamma_j ( \omega_m) \left(\frac{2}{\alpha_0 }\right)^j  \sqrt{ \frac{\alpha_0 + \alpha_2 }{2 \pi}}  \int_{-\infty}^\infty d q_3 \, ( r - q_3)^{2( \omega_m-j)} \exp \left\{ -\half ( \alpha_0 + \alpha_2 ) q_3^2 \right\}  \ . \label{xitilde0def1}
\end{align}
Thus, the integral that we must evaluate now is
 \begin{align} \label{iomegaint}
 I_{\omega } ( r ) & \equiv    \sqrt{ \frac{\alpha_0 + \alpha_2 }{2 \pi}}   \int_{-\infty}^\infty d q_3 \, ( r - q_3)^{2  \omega } \, \exp \left\{ -\half ( \alpha_0 + \alpha_2 ) q_3^2 \right\} \ . 
 \end{align}
This integral can be done analytically in terms of the confluent hypergeometric function of the second kind (also known as the hypergeometric $U$ function) to give
\be \label{exacti}
 I_{\omega} ( r )   = \lim_{\epsilon \rightarrow 0^+} \frac{  2^{\omega} e^{i \pi \omega} }{ ( \alpha_0 (r ) + \alpha_2 (r) )^{\omega}} \, U \left(-\omega , \frac{1}{2} , - \frac{r^2(1 + i \epsilon)^2}{2} ( \alpha_0 (r) + \alpha_2(r)) \right) \quad  \text{for} \quad \im \, \omega \geq 0  \ ,
 \ee
and $I_\omega(r) = I_{\omega^*} ( r )^{*}$ for $\im \, \omega < 0$.  Instead of taking the limit, in practice we simply take $\epsilon$ to be a small positive real number.\footnote{The hypergeometric $U$ function $U(a,b,z)$ is a solution of 
\be
z \frac{ d^2 U}{d z^2} +(b-z) \frac{d U}{d z } -a U = 0 \ ,
\ee
with the limiting form $U(a,b,z) \sim z^{-a}$ for $|z|\rightarrow \infty$, and has an integral representation
\be
U(a,b,z) = \frac{1}{\Gamma(a)} \int_0^\infty d t \, e^{-zt} t^{a-1} (1+t)^{b-a-1} \ . 
\ee
The small positive real number $\epsilon$ must be included in \eqn{exacti}, even when evaluating numerically.  This is because the power $(r - q_3)^{2 \omega}$ in \eqn{iomegaint} has a branch cut on the real axis $(-\infty , 0]$ when $\im \, \omega \neq 0$: the $\epsilon$ chooses the correct side of the branch cut.  In practice, we take $\epsilon = 10^{-8}$ in our computations.  
 }  Letting $\tilde \xi^{(0),N_\Gamma}_{\omega_m}$ denote the expression in \eqn{xitilde0def1} summed from $j = 0$ to $j=N_{\Gamma}$, we have
\begin{align} \label{xitildengamma}
\tilde \xi^{(0),N_\Gamma}_{\omega_m} ( r ) =   \sum_{j=0}^{N_\Gamma} \gamma_j ( \omega_m ) \left( \frac{2}{\alpha_0(r)} \right)^j   I_{\omega_m-j} ( r ) \ . 
\end{align}
This brings us to the final expression for the tree-level resummation, expanded up to order $N_\Gamma$, 
\be
\xi(r)\big|^{N_\Gamma}_0 =  \sum_m c_m \tilde M_{11}( \nu_m ) \tilde \xi^{(0),N_\Gamma}_{\omega_m} (r) \ .
\ee
In practice, as we will show, $N_{\Gamma} = 1$ is sufficient to have better than about $0.2\%$ accuracy.

%%%%%%%%%%%%%
%
%
%
\subsection{One-loop resummation}
 The expression for the resummation at one loop is (see \appref{irresumexpsec} for details), 
 \be \label{oneloopresum}
 \xi ( r ) \big|_1 = \int d^3 q \left\{ R_1 ( \rvec - \qvec , \rvec ) \xi_{11}^{\rm E} ( q ) + R_0 ( \rvec - \qvec , \rvec ) \xi_1^{\rm E} ( q ) \right\} \ .
 \ee
The approach to computing each term above is different, so we will look at each piece separately.  

We start with the second term, the $R_0 \xi^{\rm E}_1$ term, which is most like the tree-level case.  Here, we must compute
 \be
  \xi_\sigma ( r ) \big|_1^{(0,1)} \equiv \int d^3 q \,  R_0 ( \rvec - \qvec , \rvec ) \xi_\sigma^{\rm E} ( q ) 
  \ee
 where $\sigma \in \{13,22\}$; we will look at the counterterm next.  The integral to be done here is exactly of the form \eqn{tildexizero}, but with a different value of $\omega_m$.  This leads us to immediately find 
  \be \label{aoneloopeq}
   \xi_\sigma ( r ) \big|_1^{(0,1)} =   \sum_{m_1 , m_2} c_{m_1} c_{m_2} \tilde M_\sigma ( \nu_1 , \nut) \tilde \xi^{(0)}_{\omega_{1} + \omega_2} ( r ) \ .
 \ee
The counterterm is equally as simple; it is the tree-level resummation, but with $\omega_m \rightarrow \omega_m -1$, i.e.
\be \label{resumct}
\xi^\text{ct}_{13}(r) \big|^{(0,1)}_{1} \equiv \int d^3 q \,  R_0 ( \rvec - \qvec , \rvec ) \xi_{13}^{\rm ct, E} ( q ) 
 = - 4 \pi c_s^2 \, \sum_m c_m \tilde M_{11}( \nu_m -1)   \tilde \xi^{(0)}_{\omega_m -1} ( r )  \ .
\ee

Next we move on to the $R_1 \xi_0^{\rm E}$ term in \eqn{oneloopresum}.  We have 
\be \label{reexpress1}
 \xi ( r ) \big|_1^{(1,0) } \equiv \int d^3 q \, R_1 ( \rvec - \qvec , \rvec ) \xi_{11}^{\rm E} ( q ) = \int d^3 q \, \left(  \frac{5}{2} R_0 ( \rvec - \qvec , \rvec ) + \partial_\lambda R_0^\lambda ( \rvec - \qvec , \rvec) \big|_{\lambda = 1} \right)  \xi_{11}^{\rm E} ( q ) \ ,
\ee
where we have used \eqn{r1eq} to replace $R_1$.  The first term above is simply the tree-level resummation, so we will move on to the second.  For this, it is useful to define 
\be \label{xitildelambda}
 \tilde \xi_{\omega_m}^\lambda ( r ) \equiv   \frac{ (2 \pi)^{-3/2}}{\sqrt{| A ( \rvec ) |}} \int d^3 q \, q^{2 \omega_m} \exp \left\{ - \frac{\lambda}{2} (\rvec - \qvec)^i A_{ij}^{-1} ( \rvec ) (\rvec - \qvec)^j \right\}  \ .
\ee
and
\be \label{lambdaderiv}
\tilde \xi^{(1)}_{\omega_m } ( r )  \equiv \partial_\lambda  \tilde \xi_{\omega_m}^\lambda ( r ) \big|_{\lambda = 1} \ , 
\ee
in terms of which the second term in \eqn{reexpress1} can be written 
\be \label{random1}
 \int d^3 q \, \partial_\lambda R_0^\lambda ( \rvec - \qvec , \rvec )  \big|_{\lambda = 1} \xi_{11}^{\rm E}(q) = \sum_m c_m \tilde M_{11}( \nu_m ) \tilde \xi^{(1)}_{\omega_m} ( r )  \ . 
\ee
Now, one can do the integrals over $q_1$ and $q_2$ in \eqn{xitildelambda} analytically as before.  After doing that, expanding the $\Gamma$ function as in \eqn{gammaexp1}, and taking the derivative with respect to $\lambda$, we have
\begin{align}  
\tilde \xi^{(1)}_{\omega_m} ( r ) & =   \sum_j  \gamma_j ( \omega_m ) \left( \frac{2}{\alpha_0 ( r )} \right)^j \Bigg( -  (j+1)  I_{\omega_m - j } ( r )   +  J_{\omega_m-j} ( r )   \Bigg)  \label{xi1eq}   \ ,
\end{align}
where the new ingredient is 
\be \label{jint}
J_{\omega } ( r ) \equiv      \left(- \frac{ \alpha_0 + \alpha_2}{2} \right)  \sqrt{ \frac{\alpha_0 + \alpha_2 }{2 \pi}}   \int_{-\infty}^\infty d q_3 \, ( r - q_3)^{2 \omega } \,q_3^2 \, \exp \left\{ -\frac{1}{2} ( \alpha_0 + \alpha_2 ) q_3^2 \right\} \ . 
\ee
By writing $q_3^2 = r^2 - 2 r ( r - q_3) + (r-q_3)^2$, we see that the integral is written in terms of pieces which are all of the form \eqn{iomegaint}, and we can immediately write down 
\be \label{jexpression}
J_{\omega} ( r )  =\left(- \frac{ \alpha_0 ( r )  + \alpha_2 ( r ) }{2} \right)  \left(  r^2 I_{\omega} ( r ) - 2 r I_{\omega + 1/2}(r)  + I_{\omega + 1} (r)  \right) \ . 
\ee
With this in mind, the final expression for the $R_1 \xi_0^{\rm E}$ term is
\begin{align}
 \xi ( r ) \big|_1^{(1,0) } &  = \frac{5}{2}  \xi ( r ) \big|_0+  \sum_m c_m \tilde M_{11} ( \nu_m )   \tilde \xi^{(1)}_{\omega_m}  (r) \ .
\end{align}

As in \eqn{xitildengamma}, we can define $\tilde \xi^{(1),N_\Gamma}_{\omega_m} ( r )$ as the expansion of the $\Gamma$ function up to order $N_\Gamma$ for the expression of $\tilde \xi^{(1)}_{\omega_m} ( r )$ in \eqn{xi1eq}.  Then, as a final summary, the total one-loop resummed correlation function, expanded to order $N_\Gamma$ is
\begin{align}
\begin{split} \label{xioneloop}
 \xi ( r ) \big|_1^{N_\Gamma} & =  \frac{5}{2}  \xi ( r ) \big|_0^{N_\Gamma} +  \sum_m c_m \tilde M_{11} ( \nu_m )   \tilde \xi^{(1),N_\Gamma}_{\omega_m}  (r) - 4 \pi c_s^2 \, \sum_m c_m \tilde M_{11}( \nu_m -1)   \tilde \xi^{(0),N_\Gamma}_{\omega_m -1} ( r )  \\
  & \quad \quad +  \sum_{m_1 , m_2} c_{m_1} c_{m_2} \left(  \tilde M_{13} ( \nu_1 , \nut)+  \tilde M_{22} ( \nu_1 , \nut) \right) \tilde \xi^{(0),N_\Gamma}_{\omega_{1} + \omega_2} ( r )   \ . 
\end{split}
\end{align}
Thus, we have successfully given an analytic formula for the one-loop IR-resummation of the correlation function.  Next, in \secref{evalsec}, we will discuss a few different strategies to evaluate the final expression in \eqn{xioneloop}.  Then, in \secref{resultssec}, we will see how the different strategies discussed in \secref{evalsec} compare to the standard numerical integral method for the IR-resummation.

%%%%%%%%%%%%%%%
%
%
%
%
%%%%%%%%%%%%%%%%

\section{Evaluation strategies} \label{evalsec}
Although we have an analytic expression for the IR-resummed correlation function, there are some challenges to the practical evaluation of the expression in \eqn{xioneloop}.  In this section, we describe three methods for evaluating the IR-resummation so that it is as fast as possible.

%%%%%%%%%%%%%%%
%
%

\subsection{Exact evaluation} \label{exatsec}

As a practical matter, evaluating the basis functions $\tilde \xi_\omega^{(0)} (r )$ and $\tilde \xi_\omega^{(1)} (r )$ for the number of necessary $\omega$'s is too slow; in particular, the one-loop term, which has $\tilde \xi_{\omega_1 + \omega_2}^{(0)} (r )$, must be evaluated at the $\mathcal{O}(N_{\rm max}^2/4)$ values of $\omega_1 + \omega_2$.\footnote{The factor of $1/4$ comes from the fact that, first, $\tilde \xi_{\omega_1 + \omega_2}^{(0)} (r )$ is symmetric in $\omega_1$ and $\omega_2$, and second, by looking at the explicit formula for $I_\omega ( r )$ in \eqn{exacti}, that $\tilde \xi_{\omega_1^* + \omega_2^*}^{(0)} (r ) = \tilde \xi_{\omega_1 + \omega_2}^{(0)} (r )^*$.}  The bottleneck in the computation is evaluating the hypergeometric function $ U \left(-\omega , \frac{1}{2} , - \frac{r^2(1 + i \epsilon)^2}{2} ( \alpha_0(r) + \alpha_2(r)) \right)$ in \eqn{exacti} for all of the necessary values of $\omega$.  Furthermore, as written in \eqn{exacti}, the argument of the hypergeometric function depends on the cosmology through $\alpha_0(r)$ and $\alpha_2(r)$ (note that the values of $\omega_m$ do not depend on cosmology, just the interval over which we do the FFT for the initial decomposition of the power spectrum).  This means that the basis functions $\tilde \xi_\omega^{(0)} (r )$ and $\tilde \xi_\omega^{(1)} (r )$ would have to be evaluated at the relevant $\omega$ each time that we change cosmology, and this indeed is too slow to represent any improvement over existing IR-resummation techniques.  However, we can sidestep this issue by scaling out the dependence on $\alpha_0(r) + \alpha_2(r)$, thus allowing us to evaluate the hypergeometric functions once and for all in a cosmology independent way.  Then, the values can be saved in a table that can be used for any cosmology, so that the computation is made much faster.  We now describe this process.

In general, we want to obtain the correlation function at a set of points $\{ r_i\}$.  Let us imagine that we evaluate the hypergeometric functions $U \left(-\omega , \frac{1}{2} , - \frac{x^2(1 + i \epsilon)^2}{2}  \right)$, for all relevant values of $\omega$, at a fixed set of points $\{x_i \}$.  Then, in a given cosmology, we should find the inverse of the rescaling $x = r \[ \alpha_0(r) + \alpha_2 ( r ) \]^{1/2}$ that we did inside of the hypergeometric functions.  So let the function $r(x)$ be this inverse, i.e. let it satisfy $x = r(x) \[ \alpha_0 ( r(x)) + \alpha_2 ( r (x))\]^{1/2}$. Then, we have (taking $\im \, \omega >0$ in \eqn{exacti} to be concrete),
\be \label{exactscaleform}
I_\omega ( r ( x )) = 2^{\omega} e^{i \pi \omega}  ( \alpha_0 (r ( x )) + \alpha_2( r ( x ) )  )^{-\omega} \, U \left(-\omega , \frac{1}{2} , - \frac{x^2(1 + i \epsilon)^2}{2}\right) \ . 
\ee
In this way, we will obtain the correlation function at a set of points $\{ r_i = r ( x_i ) \}$.\footnote{This inversion can be easily obtained numerically by using Mathematica's FindRoot command, for example.}  This set of points depends on cosmology, but this is no worry since we will end up interpolating through the points anyway.  In this way, the hypergeometric functions only need to be computed once and then can be saved as a table and used for any cosmology.  Only the prefactors, $2^{\omega} e^{i \pi \omega}  ( \alpha_0 (r ( x )) + \alpha_2( r ( x ) )  )^{-\omega}$, need to be computed for each cosmology.  However, since they are simple functions, they evaluate very quickly.  The same approach can obviously be taken for the function $J_\omega ( r )$ in \eqn{jexpression}, which indeed is written directly in terms of $I_\omega ( r )$.  In \appref{rrapp}, we present a recursion relation for the hypergeometric function $U(a,b;z)$ that appears in \eqn{exactscaleform} which relates the $I_{\omega - j}$ for integer $j$ and can help limit the number of independent tables which must be computed.

%%%%%%%%%%%%%%%
%
%
%
%
%%%%%%%%%%%%%%

\subsection{Saddle-point approximation} \label{saddlesec}

Because the naive evaluation of the hypergeometric function $U$ in \eqn{exacti} is too slow for the number of $\omega$'s that we need for the one-loop expression in \eqn{xioneloop} (though the improvements we gave above make it actually feasible), we could imagine not using the explicit form of $I_\omega ( r ) $ in \eqn{exacti}, but instead approximating the integral in \eqn{iomegaint}.  Indeed, the form of the integral lends itself to the saddle-point approximation, especially because the expression for the saddle point can be found exactly.  We discuss this approach now.

First of all, we remind the reader that for the one-loop computation, the only integral that we have to do is for $I_\omega(r)$ in \eqn{iomegaint} because the one-loop integral for $J_\omega ( r )$ (and indeed, as we will see in \appref{higherloopssec}, for all loops) can be written in terms of $I_\omega ( r )$, see \eqn{jexpression}.  To simplify our computation, we define $x \equiv  r \sqrt{ \alpha_0(r) + \alpha_2 ( r ) }$ and $y \equiv q_3 \sqrt{ \alpha_0(r) + \alpha_2 ( r ) }$ in \eqn{iomegaint} to get 
\be \label{anintegral}
I_\omega ( r  ) =  \frac{ ( \alpha_0 ( r ) + \alpha_2 ( r ) )^{-\omega}}{\sqrt{2 \pi}} \int_{-\infty}^{\infty} d  y \, ( x - y )^{2 \omega} e^{- \half y^2} =  \frac{ ( \alpha_0 ( r ) + \alpha_2 ( r ) )^{-\omega}}{\sqrt{2 \pi}} \int_{-\infty}^{\infty} d  y \, e^{ - g_\omega ( y ) } 
\ee
where we defined $g_\omega ( y ) \equiv \half y^2 - 2 \omega \log ( x - y)$.  Because this function is so simple, we will be able to find an explicit expression for the saddle point.  To find the saddle point, we look for solutions to $g_\omega ' ( y ) = 0$, where we now consider $y$ to be a complex number.  To separate $g_\omega ( y )$ into its real and imaginary parts, it is helpful to define $ y = x- \rho e^{i \theta}$ (with $\rho \geq 0$ and $0\leq \theta <  2 \pi$), and let $\omega = \omega_r + i \omega_i$, so that we have 
\begin{align}
\begin{split}
g_\omega ( \rho , \theta ) & = \frac{x^2}{2} + 2 \theta \omega_i - x \rho \cos \theta + \half \rho^2 \left( \cos^2 \theta  - \sin^2 \theta \right)- 2 \omega_r \log \rho  \\
& \quad \quad + i \left(  - 2 \theta \omega_r - 2 \omega_i \log \rho - x \rho \sin \theta + \rho^2 \cos \theta \sin \theta    \right) \ \ .
\end{split}
\end{align}
Then, defining the real functions $u_\omega$ and $v_\omega$ as the real and imaginary parts of $g_\omega$, i.e. $g_\omega ( \rho , \theta ) = u_\omega ( \rho , \theta) + i v_\omega ( \rho , \theta)$, the condition that the complex derivative is zero is
 \be \label{saddleeq}
 \cos \theta \,  \frac{ \partial u_{\omega}}{\partial \rho}  - \frac{\sin \theta}{\rho}  \frac{ \partial u_{\omega} }{\partial \theta} = 0 \ , \hspace{.5in}  \cos \theta \, \frac{ \partial v_{\omega}}{\partial \rho}  - \frac{\sin \theta}{\rho} \frac{ \partial v_{\omega}}{\partial \theta}  = 0  \ .
\ee
This set of equations generates multiple solutions which can be found analytically with, for example, Mathematica.  Generically, there will be two viable solutions with  $\rho \geq 0$ and $0 \leq \theta  <  2 \pi$: one of them will have $\rho \approx x$ and one of them will have $\rho \ll x$.  As expected, one can check that the solution with $y \ll x$, i.e. with $\rho \approx x$, is the dominant saddle, since it is larger by at least about a factor of $e^{\half x^2}$ than the sub-dominant saddle, and the smallest we have is $x\sim 5$ (see \figref{alphaplot}) .  In the rest of this section, we will only include the dominant saddle, which we label $( \rho_s ,  \theta_s)$, although it is trivial to include the subdominant saddle if one wishes (which however appears to be completely negligible).  Going back to the variable $y = x - \rho e^{i \theta}$, the saddle point depends on the value of $\omega$ and the point at which we want to evaluate the correlation function $x(r)$, so we will label the solution as $y_s [ x , \omega]$.  The explicit solution is, for $\omega_i > 0$, 
\begin{align}
\begin{split}
\rho_s  = \half \left(   x_1^2 + x_2^2 + x^2 \right)^{1/2}  \ ,  \quad \text{and}  \quad \cos \theta_s  = \frac{x_1^2 x_2^2 + 4 x^4 + x^2 \left( 3 x_1^2 + 4 x_2^2 + 16 \omega_r \right)}{2 x \left(   x_1^2 + x_2^2 + x^2 \right)^{3/2}  } \ ,
\end{split}
\end{align}
where $x_1^2 \equiv \sqrt{  2 x^2 ( x_2^2 + x^2 + 8 \omega_r ) }$, $x_2^2 \equiv \sqrt{ x^4 + 16 x^2 \omega_r + 64 ( \omega_i^2 + \omega_r^2) }$, and $\sin \theta_s = + \sqrt{ 1 - \cos^2 \theta_s}$.   Using the explicit equations in \eqn{saddleeq}, one can easily check that  $y_s [ x , \omega^*] = y_s [ x , \omega]^*$, so that when $\omega_i<0$, the solution is the same as above, but with $\sin \theta_s = - \sqrt{ 1 - \cos^2 \theta_s}$.

With the saddle point in hand, we can expand $g_\omega ( y)$ around the saddle and shift the integration variable to $\Delta y = y - y_s$ in the integral in \eqn{anintegral} to get
\begin{align} \label{expapprox}
\begin{split}
\int_{-\infty}^{\infty} d  y \, e^{ - g_\omega ( y ) }  & = e^{ - g_\omega ( y_s ) } \int_{-\infty}^\infty d \Delta y \, e^{- \half g_\omega '' ( y_s ) \Delta y^2  - \frac{1}{3!} g_{\omega}^{(3)} ( y_s ) \Delta y^3 - \frac{1}{4!} g_\omega^{(4)} ( y_s ) \Delta y^4 + \cdots } \\
& =   e^{ - g_\omega ( y_s ) } \int_{-\infty}^\infty d \Delta y \, e^{- \half g_\omega '' ( y_s ) \Delta y^2  } \left( 1  - \frac{1}{4!} g_\omega^{(4)} ( y_s ) \Delta y^4 + \cdots  \right)  \ ,
\end{split}
\end{align}
where we have omitted the $\Delta y^3$ term in the last line because it integrates to zero here, and $g_\omega^{(n)}  \equiv \partial^n  g_{\omega} / \partial y^n $.  Now, one can keep as many terms as necessary in \eqn{expapprox}, but we will see in \secref{resultssec} that for practical purposes we only need the first term, which is simply a Gaussian integral.  The other terms in the series are higher moments of the Gaussian, and all of the integrals can be done analytically.  

Before moving on, let us comment on the size of the higher order corrections to the saddle-point approximation.  To find it, we need to estimate the size of the $g^{(4)}_\omega ( y_s)$ term in \eqn{expapprox}.  The size of the region of integration is set by the overall exponential, so we can approximate $\Delta y^2 \approx 2 / g_\omega '' ( y_s)$.  Next, at the order of magnitude level, we have $g_\omega '' ( y_s ) \approx 1 - 2 \omega / x^2$, and $g^{(4)}_\omega ( y_s ) \approx 12 \omega / x^4$, where we have used $y_s \ll x$.  Now, the typical size of $x$ near the BAO peak is between 10 and 20 (see \figref{alphaplot}), and a conservative value for the largest value of $\omega$ that one has to use is $\im \, \omega \approx 20$ (using $\nu = -0.2$, $N_{\rm max} = 200$, $k_{\rm max} = 100$, and $k_{\rm min} = 10^{-5}$ in \eqn{cmetam}).  All of this together means that we can estimate
\be \label{saddelapprox}
\int_{-\infty}^{\infty} d  y \, e^{ - g_\omega ( y ) }  \approx e^{-g_\omega ( y_s )} \sqrt{\frac{ 2 \pi}{g_\omega '' ( y_s) }} \left( 1 + \mathcal{O}\left( \frac{2 \omega}{x^4} \right)   \right) \ .
\ee
As we will see later, without the $2 \omega / x^4$ correction, the saddle-point approximation works to better than $0.2\%$, and since this additional subleading correction is less than $2 \omega / x^4 \sim 4  \times 10^{-3}$, we are justified in dropping it.  However, as we mentioned, there is no difficulty in including the higher order corrections.  We can also compare this expansion to the one that we did in \eqn{gammaexp1} for the $\Gamma$ function in order to see which is the leading correction.  There, the first correction came at order $ \omega / z$ for $z \approx \alpha_0 ( r ) r^2 / 2$, which is roughly the same order as the $\mathcal{O}(\omega / x^2)$ correction in $g_\omega '' ( y_s ) \approx 1 - 2 \omega / x^2$ (recall that $x^2/2 \equiv r^2 \left( \alpha_0 ( r ) + \alpha_2(r) \right)/2$, and so $z$ is not quite equal to $x^2 /2$).  Thus, if we use the full $g_\omega '' ( y_s )$ in \eqn{saddelapprox}, we should use the $\Gamma$ function expansion to first order; if we use the next order saddle-point approximation in \eqn{saddelapprox}, we should also go to the next order in the $\Gamma$ function expansion.  Notice, however, that the expansion of the $\Gamma$ function is not quite the same as the saddle-point expansion.  The $\Gamma$ function is an expansion in powers of $\omega / z$, while the saddle-point approximation is an expansion in $\omega / x $ times powers of $1/x$, i.e. there is simply a linear power of $\omega$, which is the potentially large parameter.  This means that the $\Gamma$ function expansion is expected to be more important in general.

Thus, for the order that we work in this paper, we only need up to $g_\omega ''$, which is given by 
\be
g_\omega '' ( \rho , \theta ) = 1+ \rho^{-2} \left(  2 \omega_r \cos 2 \theta + 2 \omega_i \sin 2 \theta + 2 i ( \omega_i \cos 2 \theta - \omega_r \sin 2 \theta) \right) \ . 
\ee
In summary, the expression for the first-order saddle-point approximation is
\be
I_{\omega} ( r ) \approx \left( \alpha_0 ( r ) + \alpha_2 ( r ) \right)^{- \omega} \frac{e^{-g_\omega ( y_s )}}{\sqrt{ g_\omega '' ( y_s) }}  = \left( r - \frac{y_s}{\sqrt{\alpha_0(r) + \alpha_2 (r) }} \right)^{2 \omega} \frac{e^{- \half y_s^2} }{\sqrt{ g_\omega '' ( y_s) }} \ ,
\ee
where we must evaluate the above at $y_s = y_s \left[ r \sqrt{ \alpha_0 ( r )  + \alpha_2 ( r ) } , \omega \right]$.

%%%%%%%%%%%%%%%%
%
%
%
%%%%%%%%%%%%%%%

\subsection{Fixed-displacements approximation} \label{fixeddispsec}

In this subsection, we describe a strategy for evaluating any implementation of the IR-resummation presented in \cite{Senatore:2014via}, although it will be particularly useful when combined with the two methods described above.  The main idea is that in any cosmology reasonably close to $\Lambda$CDM, the displacements, which enter through the functions $\alpha_0(r)$ and $\alpha_2(r)$, will only be slightly different.  This fact, combined with the controlled expansion of \cite{Senatore:2014via}, means that if one were to use the displacements of a slightly different cosmology in the IR-resummation, the error with respect to the true cosmology will decrease at each loop order.  We will be more quantitative about this statement below, but first we will discuss the implications.\footnote{In particular, we will compare the following two cosmologies in this section: C1 which has cosmological parameters $\{\Omega_m,\Omega_b,\Omega_\Lambda,h, n_s, \sigma_8\} = \{0.295, 0.0468, 0.705, 0.688, 0.9676, 0.835\}$, and C2 which has cosmological parameters  $\{\Omega_m,\Omega_b,\Omega_\Lambda,h, n_s, \sigma_8\} = \{0.295, 0.04914, 0.705, 0.688, 0.9676, 0.8224\}$. We have changed $\Omega_b$ by about $4.8\%$, which in turn changes $\sigma_8$ by about $1.5\%$.  The Planck data allows for approximately a $1.6\%$ deviation of $\Omega_b$ and a $1.1\%$ deviation of $\sigma_8$ from the central values at $68\%$ confidence level \cite{Planck:2015xua}.  We have also checked that for other deviations, for example changing $A_s$, the primordial scalar amplitude, by $\approx 2 \sigma$, the results of this section are not practically changed.    \label{2cosmos} }

\begin{figure}[htb!]
 \includegraphics[width=7.5cm]{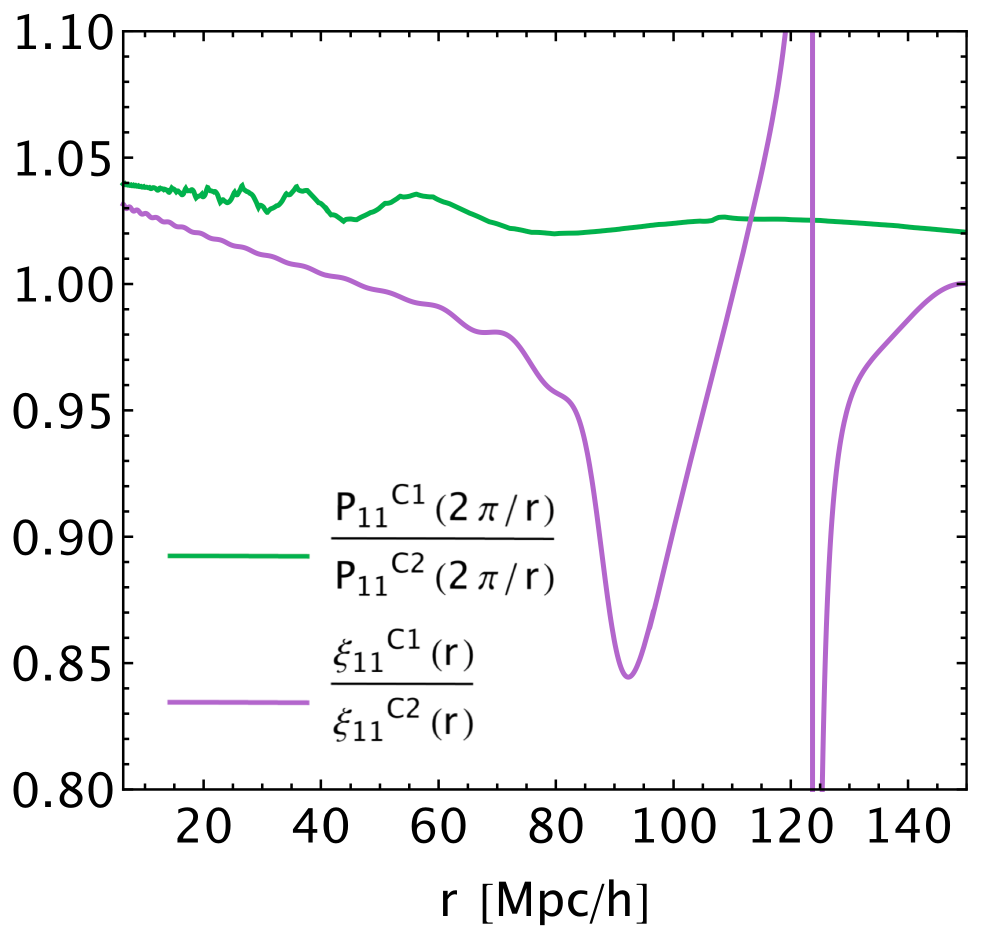}  \quad  \includegraphics[width=8.1cm]{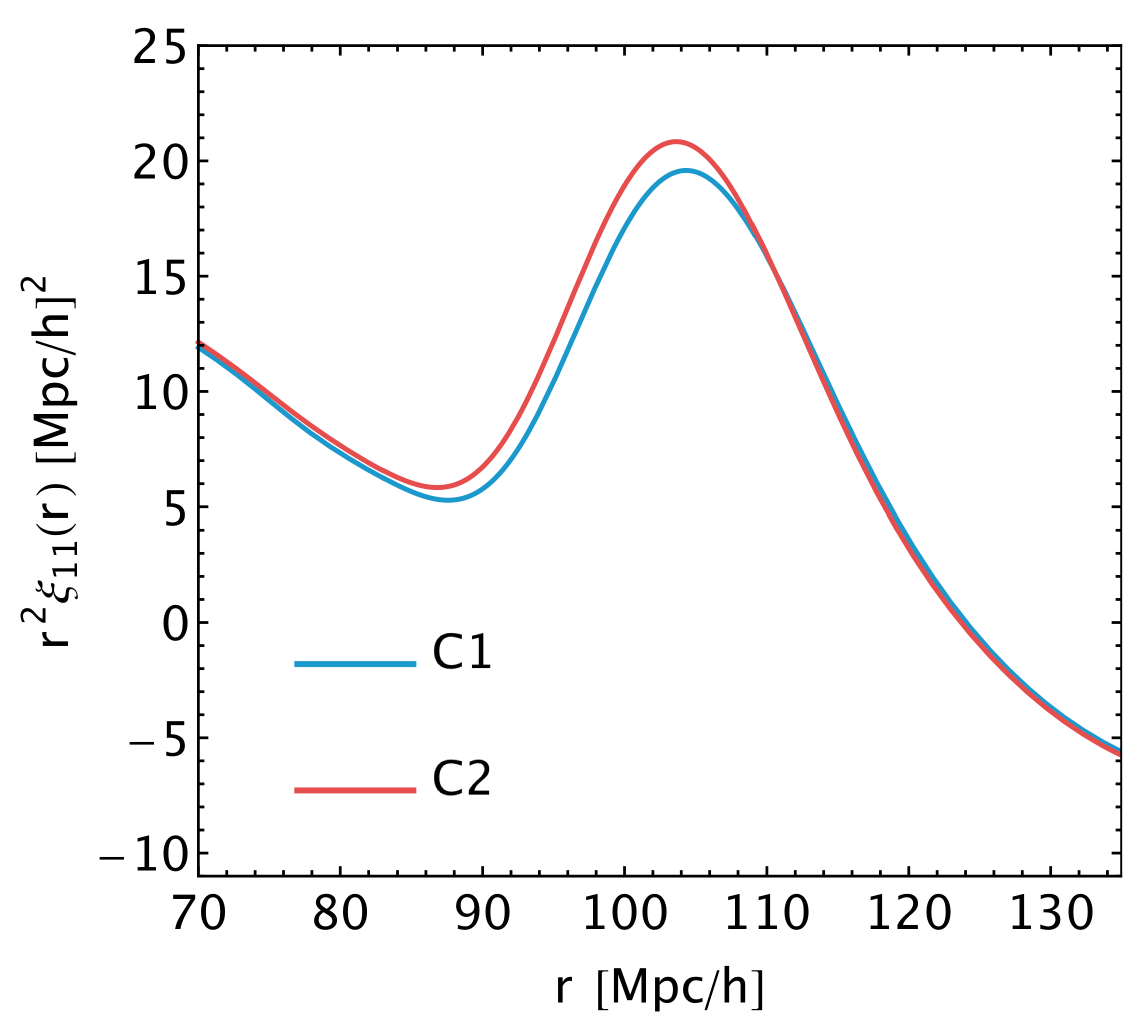}
\caption{Here we show the difference between the two cosmologies used in this paper, C1 and C2, described in \footnoteref{2cosmos}, by plotting both the linear power spectra and the linear correlation functions.  The cosmologies differ by about $4.8\%$ in $\Omega_b$ and $1.5 \%$ in $\sigma_8$, both deviations which are greater than those allowed by the $68\%$ confidence limits of Planck \cite{Planck:2015xua}.  Notice that while the power spectra are different by about $5\%$, as expected from the change in $\Omega_b$, this causes an almost $15\%$ change in the BAO peak.   We note that the correlation function goes through zero near $r \approx 120 \unitsr$, which is why the purple curve diverges above.    } \label{linearcompplot}
\end{figure}

The main benefit of the above realization is that when scanning over cosmologies to do parameter fits, one can use the displacements of a single fiducial cosmology for the IR-resummation of each separate cosmology.  This simplifies and speeds up the IR-resummation in two ways.  First, one does not need to compute the functions $\alpha_0(r)$ and $\alpha_2(r)$ for each cosmology, but instead can use the functions of a fiducial cosmology $\alpha^{\rm fid.}_0 ( r )$ and $\alpha^{\rm fid.}_2 ( r)$, for example using the Planck best fit parameters.  Secondly, and more importantly, all of the parts of the IR-resummation which depend on the displacements will then be fixed and not have to be recomputed for each cosmology.  To see this advantage more specifically, we focus on the exact resummation of \secref{exatsec}.   We mentioned in that section that one can pre-compute the hypergeometric $U$ function in \eqn{exactscaleform} and factor out all of the cosmology dependence so that one only has to compute simple prefactors for each cosmology.  However, if one fixes the displacements to some fiducial value, then the entire function $I_{\omega} ( r )$ is cosmology independent.  This means that in \eqn{xioneloop}, the functions $\tilde \xi^{(0)}_{\omega_m}(r)$ and $\tilde \xi^{(1)}_{\omega_m} (r)$ are also cosmology independent, and the sums in \eqn{xioneloop} are simple matrix multiplications where the only cosmology dependence is through the linear power spectrum coefficients $c_m$.

\begin{figure}[htb!]
\hspace{.7in} \includegraphics[width=11cm]{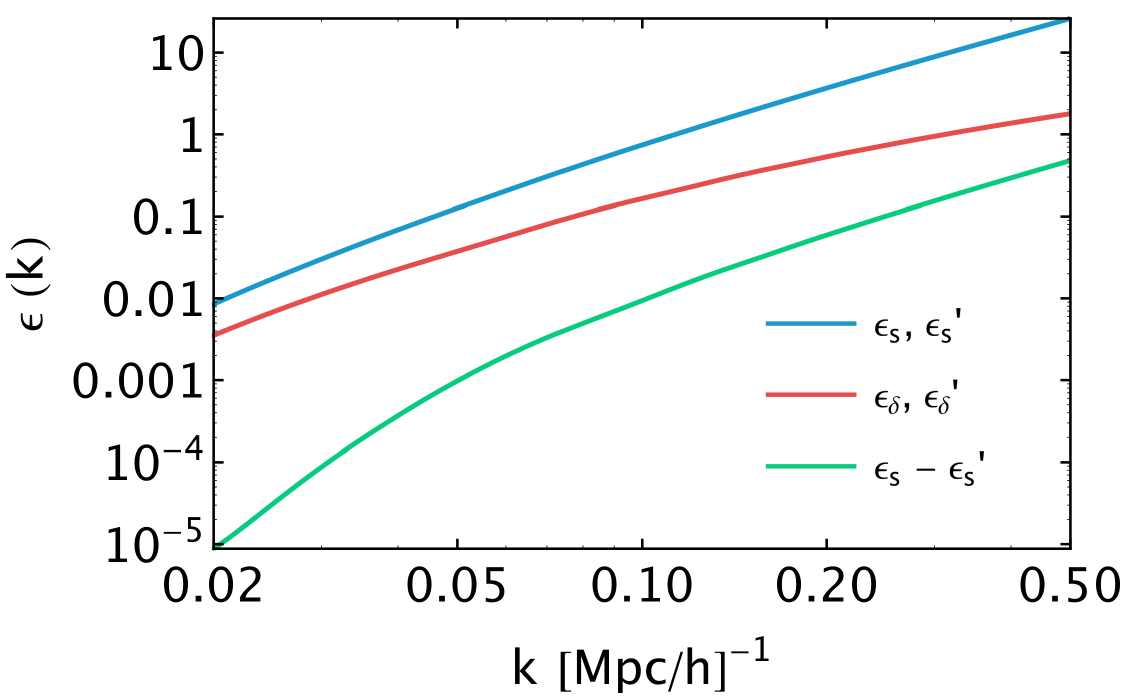}
\caption{ In this plot, we show the expansion parameters $\epsilon_{s <}$ and $\epsilon_{\delta <}$ for the cosmologies C1 and C2 described in \footnoteref{2cosmos}.  Here, the unprimed quantities come from C1 and the primed quantities come from C2.  The blue curve is $\epsilon_{s<}$ and $\epsilon_{s<}'$ (they are indistinguishable in this plot), the red curve is $\epsilon_{\delta <}$ and $\epsilon_{\delta <}'$ (they are also indistinguishable), and the green curve is the difference $\epsilon_{s<} - \epsilon_{s<}'$.  We see that the difference in the displacements between the two cosmologies, which have a different $\Omega_b$ by about $4.8\%$, is at the percent level for $ k \sim 0.12 \unitsk \equiv \Lambda_{\rm IR}$, which is the scale up to which we resum the IR modes.  } \label{epsilonsplot}
\end{figure}

Let us now justify this fixed-displacements approximation.  For simplicity, we will use the expressions from \cite{Senatore:2014via} for the IR-resummation in momentum space, although the conclusions, in particular about the dependence on $\epsilon_{s<}$, will be the same in real space.  In order to understand the parametric dependence on the displacements $\epsilon_{s<}$, we will consider a schematic version of the exact formulas from \cite{Senatore:2014via}, and we refer the reader to that reference for more details.  Furthermore, we will concentrate on the linear displacements, since it was shown that the next to leading order corrections are small \cite{Senatore:2017pbn}.  In \figref{linearcompplot}, we show the differences between the linear quantities of the two cosmologies C1 and C2 discussed in this section.

The IR-resummation of \cite{Senatore:2014via} defines a controlled expansion in $\epsilon_{\delta <}$ while keeping the exact dependence $e^{\epsilon_{s<}}$ of the linear displacements, which is analytic \cite{Senatore:2017pbn}.  In formulas, we can focus on $K \equiv   e^{\es} ( \ed + \ed^2 + \ed^3 + \cdots)$, which, very schematically, represents the exact answer for the power spectrum (the $\cdots$ stands for higher order terms in $\ed$, and for the rest of this section, $\epsilon_{s,\delta}$ stands for $\epsilon_{s<,\delta<}$ for brevity).  For the resummation at order $N$ in $\ed$ and to all orders in $\es$, we have
\be
K \big|_{N} = \sum_{j=1}^N M_{N-j}  \, K^{\rm E}_j
\ee
where $K_j^{\rm E}$ is the $j$-th order piece of the expansion of $K$ in $\es$ and $\ed$ (i.e. it is the Eulerian expansion, and for example, $K_0^{\rm E} = \ed$), 
\be
M_{N-j} = K_0 \cdot K_0^{-1} \big| \big|_{N-j} \ , 
\ee
$K_0 \equiv e^{\es}$, and the double bar $||_{N-j}$ means to expand up to order $N-j$ in both $\es$ and $\ed$ (in our example $K_0$ only depends on $\es$, but in general there can be a small dependence on $\ed$).  As an example, the one-loop expansion is 
\be
K \big|_1 = M_1 K^{\rm E}_0 + M_0 K^{\rm E}_1 = e^{\es} \left( 1 - \es \right) \ed  + e^{\es}  \ed  \left( \es +  \ed \right) = e^{\es}  \ed \left( 1 +  \ed \right) \ . 
\ee

We are now in a position to consider the effect of using the wrong displacements, called $\esp$, in the resummation.  Letting $K' |_N$ be the resummation when $\esp$ is used, we have, at tree level $ K' \big|_0 = \ed e^{\esp} $, 
so that 
\be
\frac{K' |_0 }{ K |_0} = 1 - \Delta \es \ , 
\ee
 to first order in perturbations, where $\Delta \es \equiv \es - \esp$, which as can be seen in \figref{epsilonsplot}, is a small parameter at $\Lambda_{\rm IR} = 0.12 \unitsk$, which is the scale of interest.  Furthermore, at one loop, we have
\be
K' \big|_1 = e^{\esp} \left( 1 - \esp \right) \ed + e^{\esp} \ed  ( \es +  \ed ) \ .
\ee
Notice that the $\es$ that appears in the last term is the true cosmology $\es$ because it comes from the Eulerian loop $K^{\rm E}_1$.  This gives
\be
\frac{K' |_1 }{K |_1} = 1 - \half \Delta \es \left( \Delta \es + 2  \ed \right) \ ,
\ee
to second order in perturbations, and similarly at two loops
\be
\frac{K' |_2 }{K |_2} = 1 - \frac{1}{6} \Delta \es \left( \Delta \es^2 + 3  \ed \Delta \es + 6  \ed^2 \right) \ ,
\ee
to third order in perturbations.  Looking at \figref{epsilonsplot}, we see that, taking $\Delta \es$ at $\Lambda_{\rm IR}$, $\Delta \es \lesssim \ed$, so that the leading corrections above are the ones with the most powers of $\ed$: in general, the correction to the $N$ loop resummation is 
\be \label{errorest}
\frac{K' |_{N}  - K |_{N}  }{K |_{N}} \sim \mathcal{O} \left( \Delta \es \, \ed^{N} \right) \ .
\ee
 This shows that the corrections become smaller and smaller because of the factor of $\ed^N$.  Let us keep in mind that the IR-resummed expression to order $N$ has a relative error of order $\ed^{N+1} $, so this mistake in \eqn{errorest} is an irrelevant mistake as long as $\Delta  \es < \ed $.\footnote{To be more precise and take into account the case when $\ed \lesssim \Delta \es$, the error goes like  \newline $(K' |_{N}  - K |_{N}  )/ K |_{N} \sim \mathcal{O}( \text{max} [\Delta \es \ed^N , \Delta \es^{N+1} / (N+1)! ] )$. }

\begin{figure}[htb!]
\hspace{.3in} \includegraphics[width=14cm]{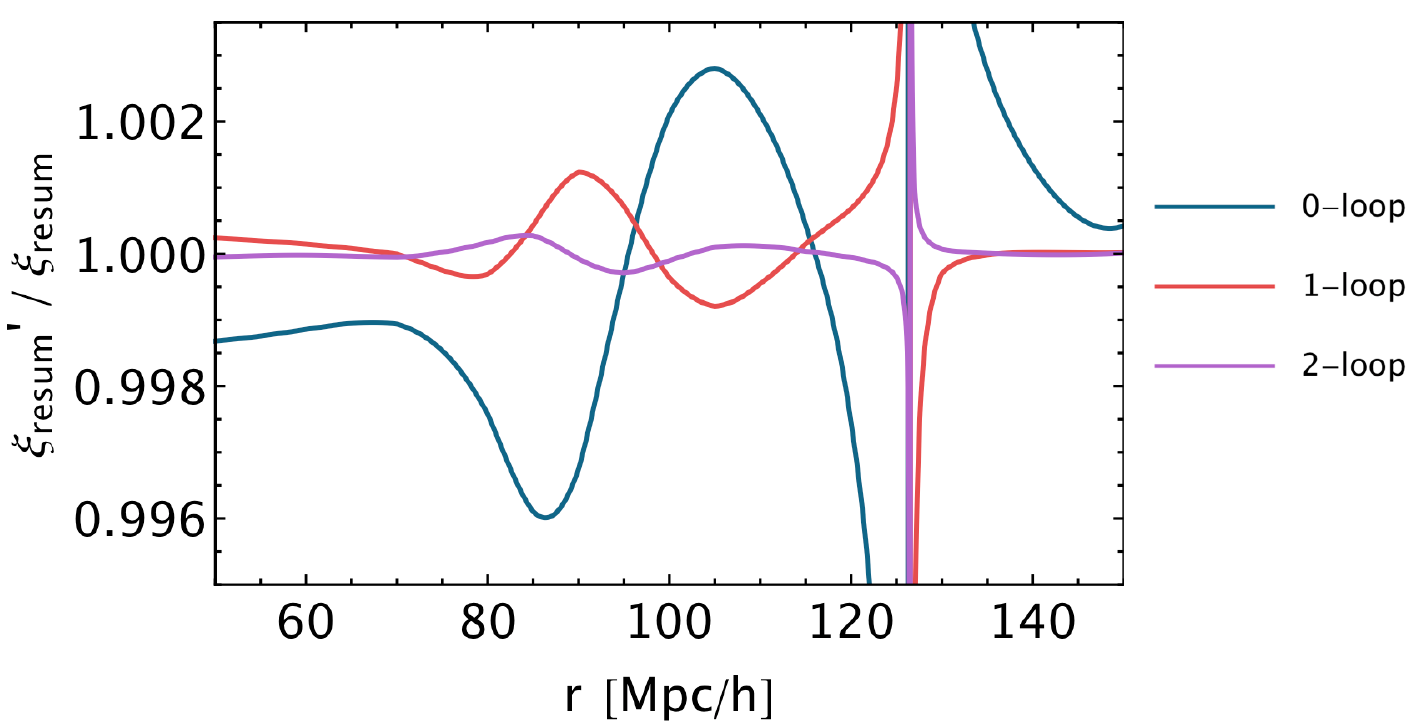}
\caption{In this plot, we show the convergence of the fixed-displacements approximation as a function of the number of loops, using the two cosmologies C1 and C2 described in \footnoteref{2cosmos}, which have values of $\Omega_b$ which are different by approximately $4.8\%$.  Here, $\xi_{\rm resum}$ is the correctly computed IR-resummed correlation function using the base cosmology C1. On the other hand, for $\xi_{\rm resum}'$, the Eulerian loops are computed in the cosmology C1, but the displacements from C2 are used in the IR-resummation.  As we see, the approximation gets better with the number of loops, in reasonable agreement with what is expected from \eqn{errorest}.}   \label{expandplot}
\end{figure}

In \figref{expandplot}, we show how this works in the computation of the IR-resummation.   In this plot, we use the two cosmologies described in \footnoteref{2cosmos}, which are the same except that $\Omega_b$ is changed by about $4.8\%$, which in turn changes $\sigma_8$ by about $1.5\%$.  The Planck data allows for approximately a $1.6\%$ deviation of $\Omega_b$ and a $1.1\%$ deviation of $\sigma_8$ from the central values at $68\%$ confidence level, so what we consider is a fairly large deviation \cite{Planck:2015xua}.  Because we want to show the result up to two loops, we use the standard numerical integral resummation of \cite{Senatore:2017pbn}, with $\Lambda_{\rm IR} = 0.12 \unitsk$.  The quantity $\xi_{\rm resum}$ is the correctly computed resummed correlation function using the base cosmology C1. In the quantity $\xi_{\rm resum}'$, the Eulerian loops are computed in the cosmology C1, but the displacements from C2 are used in the IR-resummation: \figref{epsilonsplot} shows the difference in displacements between the two cosmologies.  From \figref{expandplot}, we can see that indeed the fixed-displacements approximation gets better as the loop number increases.  We can also see that our numerical results in \figref{expandplot} are in reasonable agreement with the estimate in \eqn{errorest}. The corrections here are all sub-percent, so this method will be accurate enough to be used in upcoming data analyses.

%%%%%%%%%%%%%%%%
%
%
%
%%%%%%%%%%%%%%%

\section{Results and conclusion} \label{resultssec}

\begin{figure}[htb!]
 \includegraphics[width=15cm]{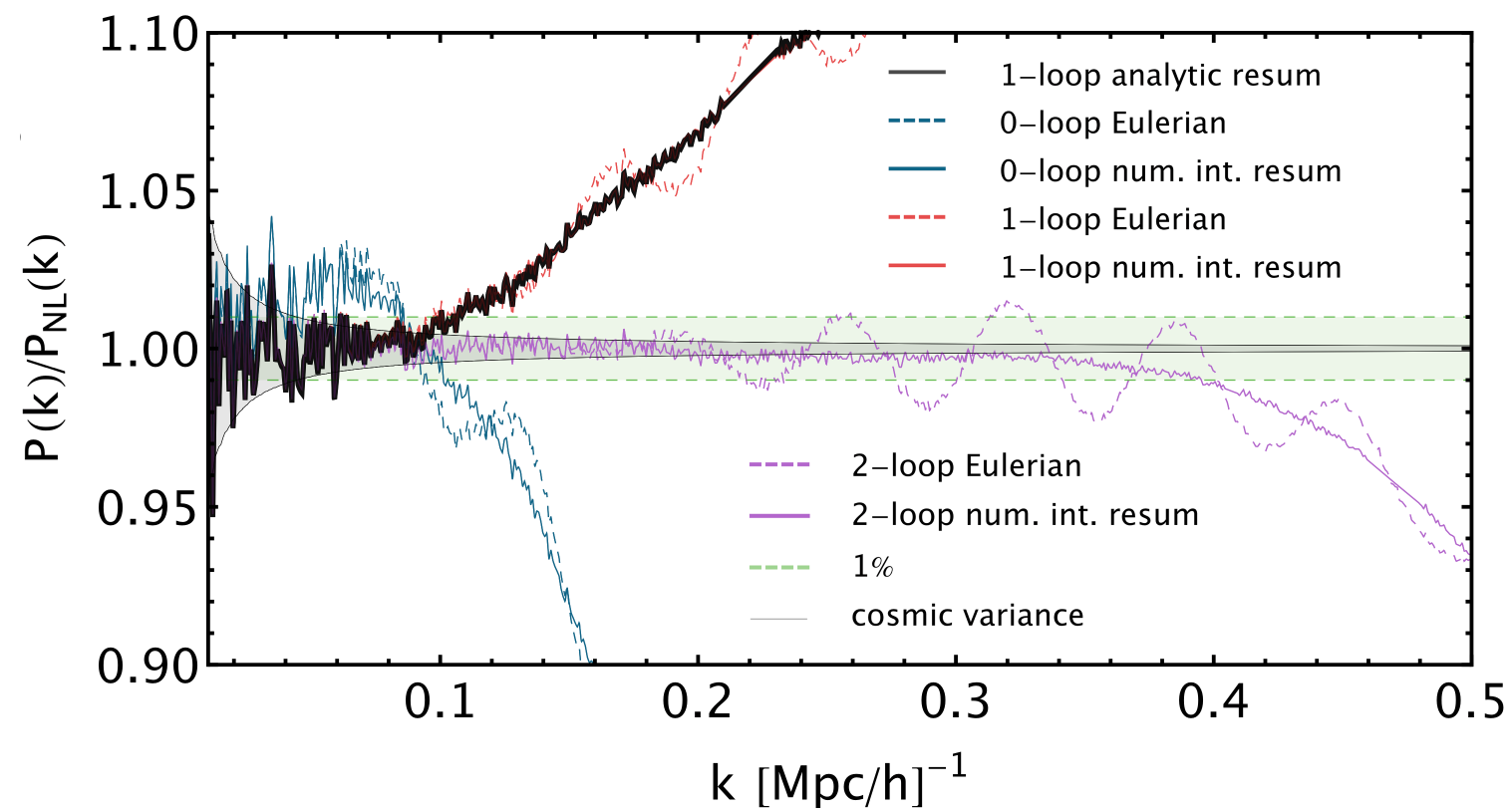} 
\caption{We show the result of the IR-resummation of the power spectrum in momentum space, compared with non-linear (NL) data from the \textsf{ds14\_a} run of the Dark Sky simulation \cite{Skillman:2014qca}.  Cosmological parameters are given in the main text, and the values for the counterterms are taken from \cite{Foreman:2015lca}.  In this plot, all of the thin resummed curves (blue, red, and purple) are evaluated with the standard numerical integral technique \cite{Senatore:2017pbn}.  The thick black curve (which is directly on top of the solid red curve) is evaluated using the exact analytic method described in this paper in \secref{exatsec}.  In momentum space, the BAO feature shows up as oscillations in the power spectrum.  The fact that the resummed curves are not wiggly in this plot means that the BAO features are being correctly taken into account. } \label{allps}
\end{figure}

In this section, we first compare our results to simulation data, and then we analyze in more numerical detail the precision and accuracy of our present developments.\footnote{A Mathematica notebook with our computations is available at the \href{http://stanford.edu/~senatore/}{EFTofLSS repository}.}    In this paper, we have presented three methods (\secref{evalsec}) for computing the analytic IR-resummation: the exact evaluation, the saddle-point approximation, and the fixed-displacements approximation.  We have already thoroughly discussed the fixed-displacements approximation in \secref{fixeddispsec}, so we will focus on the former two in this section, and would like to reiterate here that the ultimate method that we propose is a combination of the exact evaluation and the fixed-displacement approximation.  We will compare to data from the \textsf{ds14\_a}  run of the \href{http://darksky.slac.stanford.edu}{Dark Sky} simulation which evolved $10240^3$ particles in a volume of $(8\,h^{-1}\text{Gpc})^3$ with cosmological parameters  $\Omega_m = 0.295$, $\Omega_b = 0.0468$, $\Omega_\Lambda = 0.705$, $h = 0.688$, $n_s = 0.9676$, $\sigma_8 =0.835$ \cite{Skillman:2014qca}.\footnote{These are the same parameters as for the cosmology C1 in \secref{fixeddispsec}.}  For the one- and two-loop EFT expressions for the power spectrum and correlation function, we use the fitted parameters obtained in~\cite{Foreman:2015lca} which compared to the same simulation.  In particular we have, in the notation of \eqn{pexpand}, $c_s^2 ( a_0 ) = 0.0413 \left( \unitsk \right)^{-2}$.  For all IR-resummation methods, we use the cutoff $\Lambda_{\rm IR} = 0.12 \unitsk$ (defined in \eqn{a0def}).  In all legends, ``analytic'' refers to the exact evaluation in \secref{exatsec}, ``saddle'' refers to the saddle-point approximation in \secref{saddlesec}, and ``num. int.'' refers to the numerical integral method of \cite{Senatore:2017pbn}.  For the tree-level power spectrum decomposition, we use the parameters (see \eqn{cmetam}) $\nu=-0.2$, $N_{\rm max} = 200$, $k_{\rm max} = 100$, and $k_{\rm min} = 10^{-5}$, and for the one-loop decomposition, we use $\nu = -1.2$, $N_{\rm max} = 250$, $k_{\rm max} = 100$, and $k_{\rm min} = 10^{-5}$.

\begin{figure}[htb!]
 \includegraphics[width=8cm]{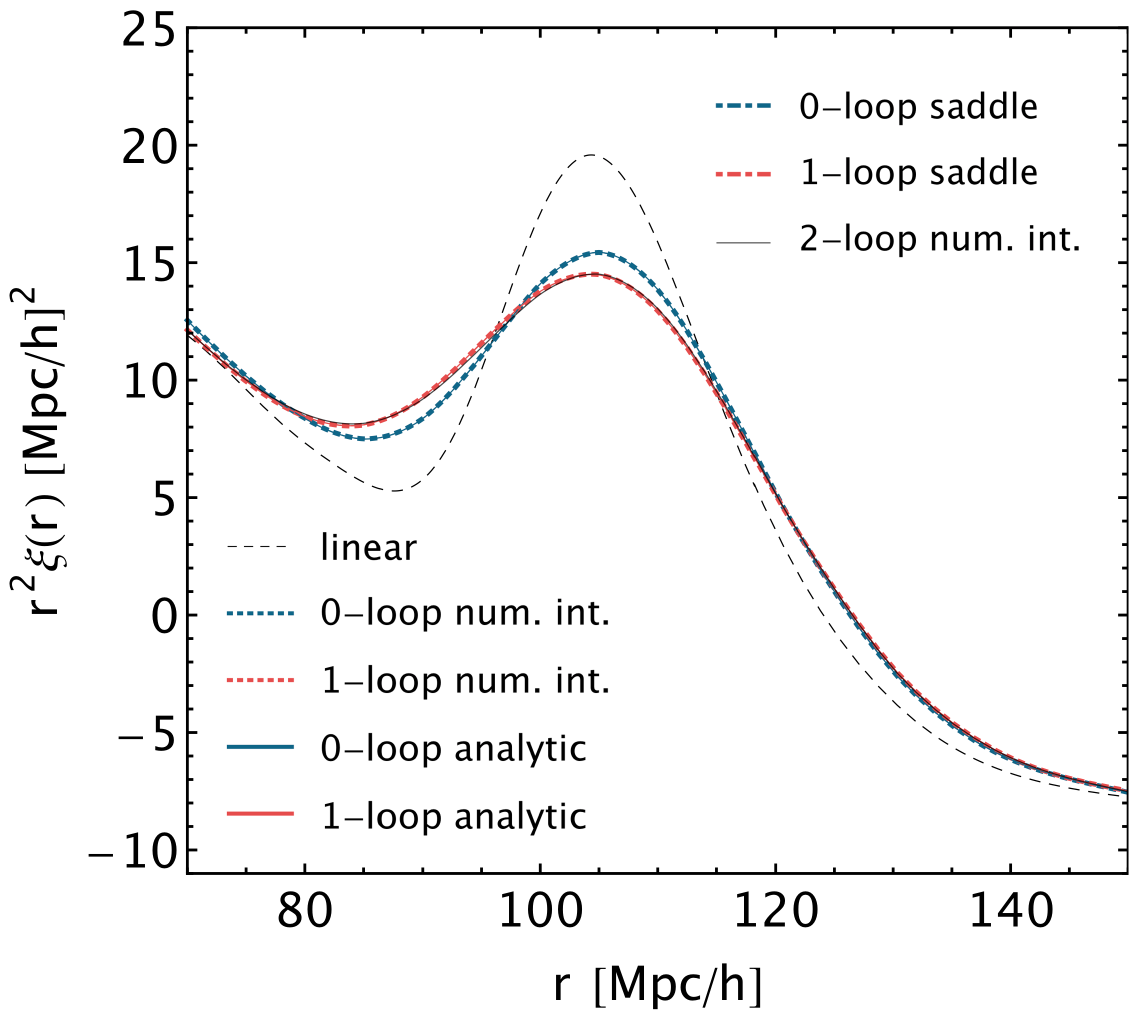}  \quad  \includegraphics[width=8cm]{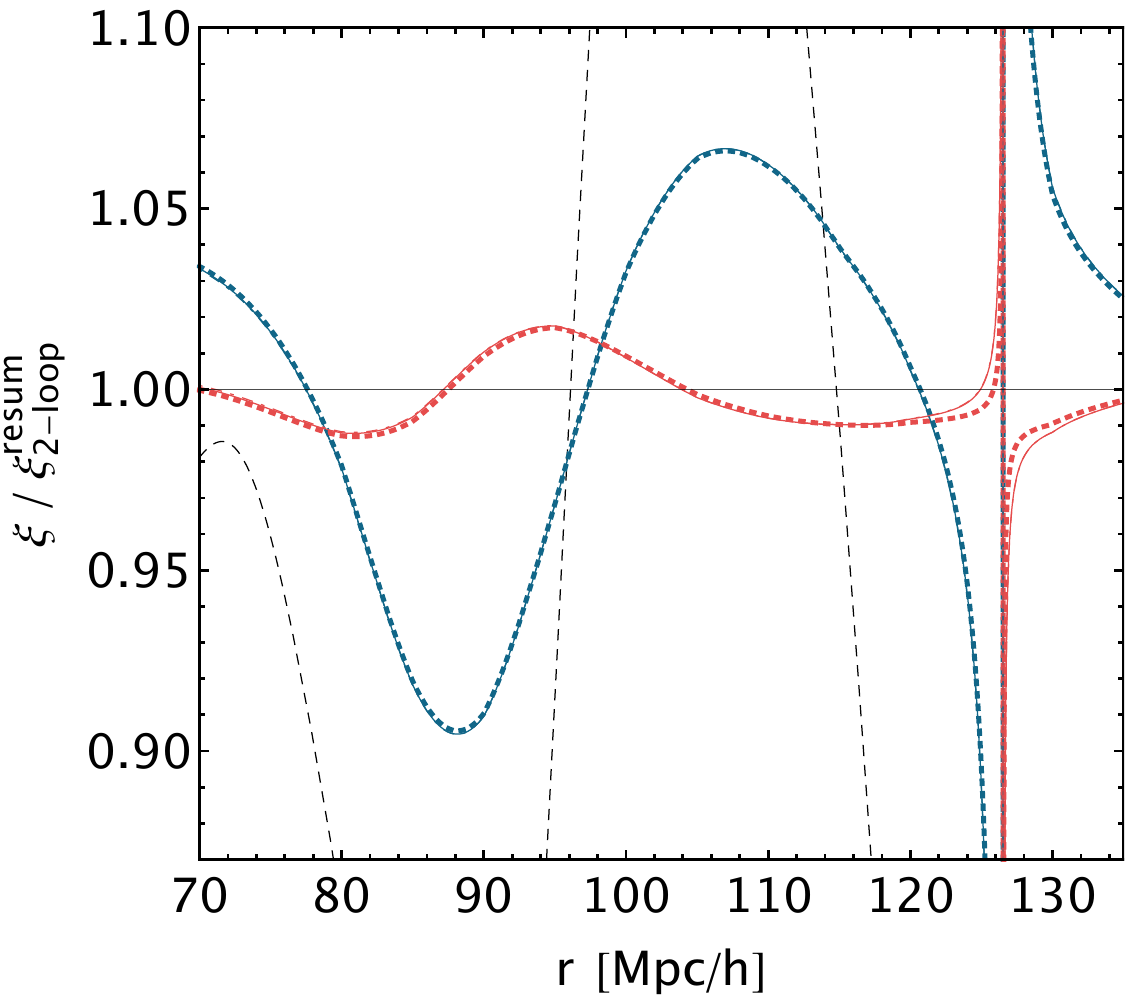}
\caption{In this figure, we compare the new exact analytic and saddle point-approximation methods of IR-resummation to the standard numerical integral (``num. int.'' in the legend) method of \cite{Senatore:2017pbn}.  In the left panel, we give an overview of the various quantities, although curves of the same color are bunched together because they are approximately equal.  In the right panel, we show some more detail by comparing the various curves directly with the two-loop numerical integral IR-resummation, which, as described at the beginning of \secref{resultssec}, we use as a proxy for the correct answer.  In the right panel, we see two things.  First, we see convergence to the correct answer, as the one-loop quantities are clearly an improvement over the tree-level quantities.  Second, we see that all three methods of computing the IR-resummation are approximately equal.  More detail is shown in \figref{compare}.  } \label{allbao}
\end{figure}

Because we are not aware of measurements of the correlation function for this simulation, we make our comparison in momentum space in \figref{allps}: the result of our one-loop, exact analytic IR-resummation (\secref{exatsec}) is shown as the black curve.  The other curves are either the Eulerian power spectra, or the IR-resummed power spectra using the numerical integral formulas of \cite{Senatore:2017pbn}.  The BAO peak of the correlation function appears as oscillations in the power spectrum, so when the curve of $P(k)/P_{\rm NL} (k)$ is wiggly, this means that the computed $P(k)$ is not correctly describing the BAO peak of the non-linear (NL) data: this is the case with all of the Eulerian curves in \figref{allps}, and this is how the poor convergence of the Eulerian correlation function curves of \figref{introplot} shows up in the power spectrum.  On the other hand, the IR-resummed curves do not show any residual oscillations.  Since we are ultimately interested in studying the BAO peak in this paper, the main point of \figref{allps} is to show that the two-loop, IR-resummed EFT power spectrum from \cite{Senatore:2017pbn} fits the the BAO oscillations very well.  Thus, since we do not have the correlation function data for this simulation, we will use the two-loop, IR-resummed EFT correlation function as a proxy for the data.  In any case, the main point of the current work is to study the accuracy of our method against the existing computational methods of \cite{Senatore:2014via,Senatore:2017pbn}.

\begin{figure}[htb!]
\quad \quad \quad \quad  \includegraphics[width=14cm]{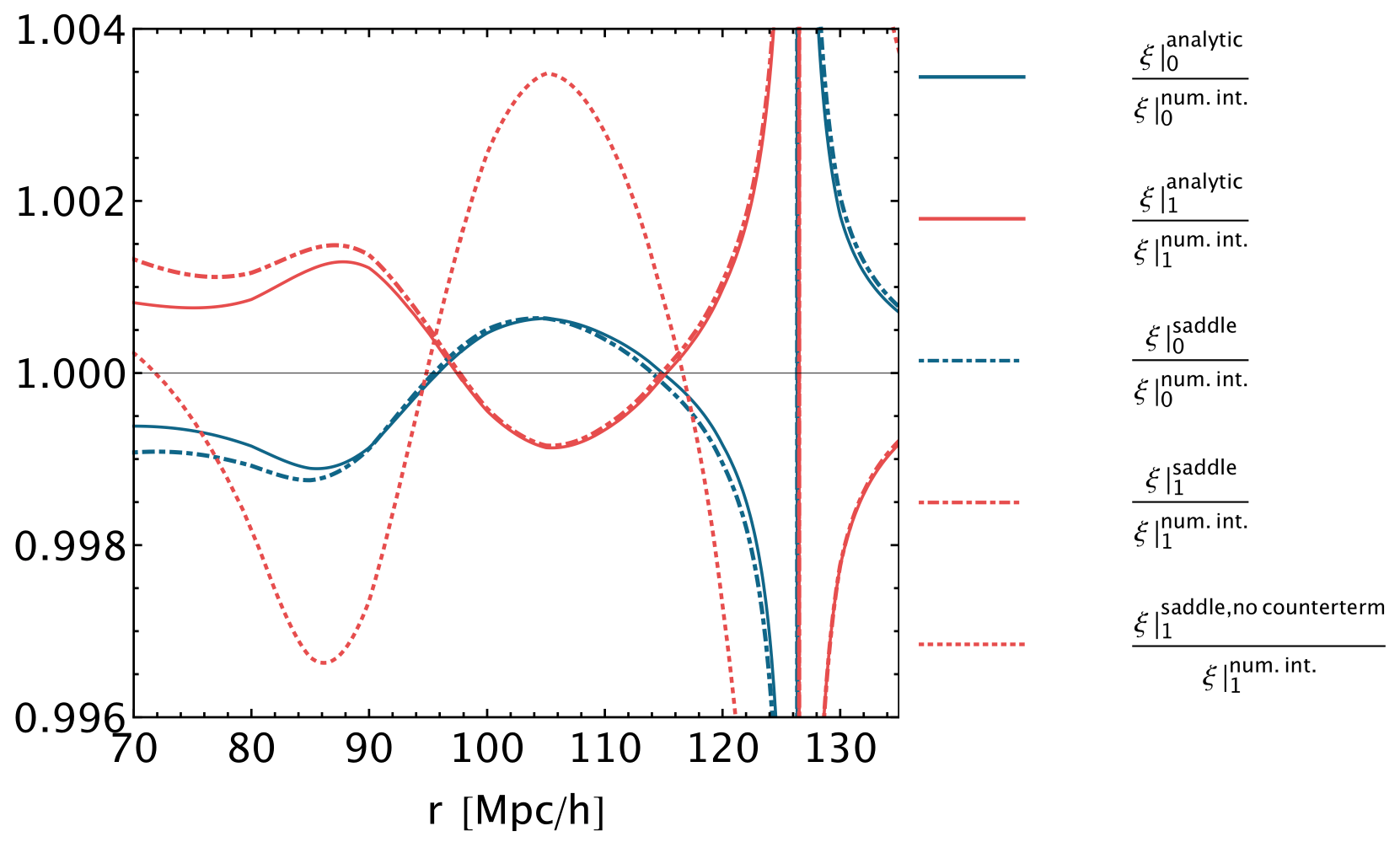}  \caption{In this figure, which is our main result, we compare our new IR-resummation methods to the standard numerical integral (``num. int.'' in the legend) IR-resummation of \cite{Senatore:2017pbn}.  We remind the reader that $\xi |_N$ means that the correlation function is resummed in $\epsilon_{s <}$, and expanded to order $N$ in $\epsilon_{\delta <}$.  We see that the two methods proposed in this work are different from the numerical integral computations by less than $0.2 \%$.  This difference is mainly due to the expansion of the $\Gamma$ function in \eqn{gammaexp1}, which also explains why the mistake is not recovered by the loop expansion.  One can easily include more terms in the $\Gamma$ function expansion, as described in the main text.  Finally, we show the effect of the counterterm, which is approximately $0.4\%$.  For this curve (dotted red), we have included the counterterm in $\xi|_1^{\rm num. int.}$, but have left it out of $\xi |_1^{\rm saddle}$.  This shows quantitatively the effect of the counterterm on the BAO scales.}  \label{compare}
\end{figure}

Now, we look directly at the correlation function.  First, in \figref{allbao}, we show an overview of two of the methods proposed in this paper (the exact analytic and saddle-point approximation methods) and the standard numerical integral method of \cite{Senatore:2017pbn}, compared with the two-loop numerical integral IR-resummation, which is our proxy for the true answer.  We can clearly see that the loop expansion is converging.  However, because the different computational methods are all so close, this plot is not very useful for showing the differences between them.  For that comparison, we look at \figref{compare}.

In \figref{compare}, which summarizes our main results, we compare our methods of IR-resummation to the standard numerical integral computation.  In particular, we compare the tree-level and one-loop resummations for both the exact analytic evaluation and the saddle-point approximation to the corresponding computations done with numerical integrals.  The difference is less than $0.2\%$ for both methods, and it does not improve with the number of loops.  This makes sense, since the expansion of the $\Gamma$ function in \eqn{gammaexp1} is common to both methods and is a systematic error which is not recovered order by order in the loop expansion.\footnote{ We can check that the order of magnitude of this mistake is what is expected.  Looking back at \eqn{gammaexp1}, the first term that we we did not include in the $\Gamma$ function expansion is of order $ \omega^2 / z^2$.  For our decomposition, we can take representative values $\im \, \omega \sim 10$ and $z \sim 170$, which gives $ \omega^2 / z^2 \sim 3.5 \times 10^{-3}$, in approximate agreement with what we see in \figref{compare}.}  This should be contrasted with the fixed-displacements approximation of \secref{fixeddispsec}, where the mistake from the displacements leads to a smaller and smaller effect as we go to higher loops.  However, this is of little importance here, since one can easily include higher orders in the $\Gamma$ function expansion, and we are already at a precision of better than $0.2\%$, which should be precise enough for any upcoming survey.  Finally, in \figref{compare}, we look at the effect of the counterterm on the BAO peak.  To get an idea, we plot the one-loop saddle-point approximation without the counterterm, divided by the one-loop numerical integral computation with the counterterm.  We see that the effect is approximately $0.4\%$.

\begin{figure}[htb!]
\quad \quad \quad \quad \includegraphics[width=14cm]{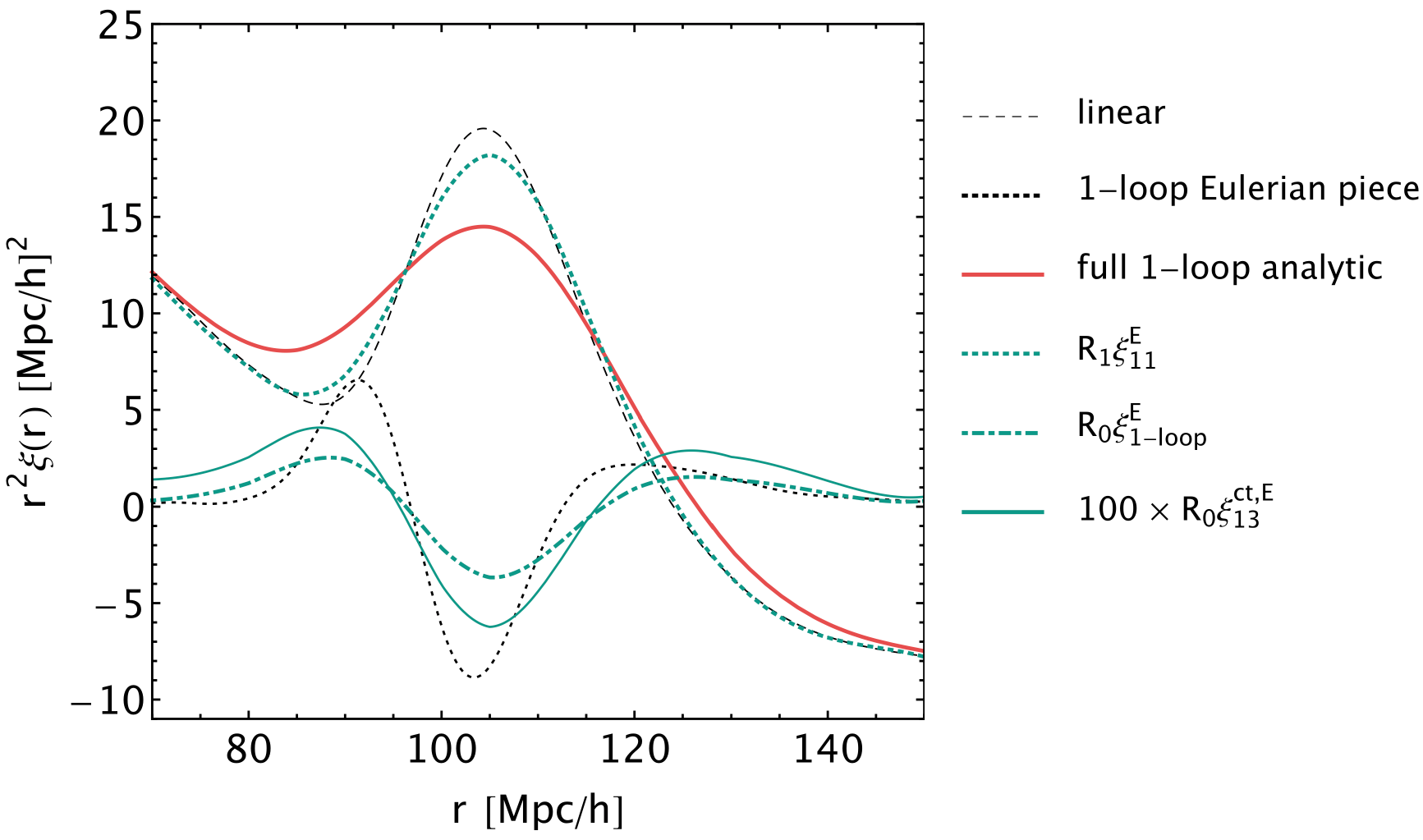}  \caption{ In this plot, we show the various contributions to the one-loop resummed correlation function.  We show the difference between the Eulerian terms (black dashed and dotted) and the corresponding resummed contributions (teal dotted and dot-dashed), as well as the counterterm contribution (solid teal).  Notice that the counterterm contribution has been scaled by a factor of $100$ to make it visible: it is much smaller than the other one-loop pieces.  For the full definitions of the $R_1 \xi_{11}^{\rm E}$ and $R_0 \xi_{1\text{-loop}}$ contributions, see \eqn{oneloopresum}.  The counterterm contribution, \eqn{resumct}, is plotted with  $c_s^2  = 0.0413 \left( \unitsk \right)^{-2}$.   } \label{pieces}
\end{figure}

To put this into perspective, in \figref{pieces} we show the various contributions to the one-loop exact analytic resummation.  We plot the full one-loop computation, along with the individual $R_1 \xi_{11}^{\rm E}$ and $R_0 \xi^{\rm E}_{1\text{-loop}}$ contributions (see \eqn{oneloopresum} for notation), and also the counterterm contribution (\eqn{resumct}, with  $c_s^2  = 0.0413 \left( \unitsk \right)^{-2}$).  Notice that the counterterm contribution has been scaled by a factor of $100$: it is about $1/100$ of the one-loop contribution.  The smallness of this contribution is to be expected since the counterterm describes UV physics, and the BAO peak is dominated by IR physics.

Now that we have seen how precise our new computations are, we can discuss the improvements in computational time.  All comparisons made here have been done in Mathematica on a laptop.  Because we expect that everything can be run much faster with a more sophisticated code, we simply report the ratio of times between different methods.  As presented in \secref{exatsec} and \secref{saddlesec}, the exact evaluation and saddle-point approximation take just about as long as the numerical integral method per output $r$ point, and so by themselves they do not represent much of an improvement.  This is observation is particularly relevant because we do not yet have the two-loop analogue of the one-loop power spectrum decomposition of \secref{decompsec}.  However, we hope that our formalism might simplify other computations, such as the IR-resummation of the bispectrum, which to date has not been implemented.  

More immediately, though, our formalism, combined with the fixed-displacements approximation of \secref{fixeddispsec} does provide a significant improvement (about a factor of $6$ to $10$ in computational time) over the standard computation.   As we remarked in \secref{fixeddispsec}, once the IR-displacements are fixed, the IR-resummation becomes simply a matrix multiplication between the cosmology dependent $c_m$ coefficients and a fixed, cosmology independent resummation matrix.  In particular, when combining the exact evaluation method of \secref{exatsec} with the fixed-displacements approximation of \secref{fixeddispsec}, we find that the error with respect to the numerical integrals is at the level of $0.3\%$, in agreement with what one would guess by combining the errors of the individual methods.

We see a few interesting extensions of this work.  First, concerning the power spectrum, we have shown in \appref{higherloopssec} that the higher loop generalization of our work should be straightforward.  Thus, once we understand the two-loop FFTLog decomposition, we should be able to apply our formalism to it without much trouble.  Next, we hope that our work can also be used to implement the IR-resummation for other cosmological observables, like the bispectrum and redshift space distortions.  In those cases, there are more numerical integrals to be done, and practical implementation is more difficult than for the correlation function (for example, it does not exist yet for the bispectrum).  As always, having an analytical formula may help with simplifying numerical computations, but we leave this exploration for future work.   

%%%%%%%%%%%%%%
%
%
%
%
%%%%%%%%%%%%

%%%%%%%%%%%%%%%%%%%%%
%
%%%%%%%%%%%%%%%%%
\section*{Acknowledgments}

The authors would like to thank J.~J.~M. Carrasco for interesting discussions related to this project.  L.~S. is partially supported by NSF award 1720397 and partially supported by Simons Foundation Origins of the Universe program (Modern Inflationary Cosmology collaboration).  M.~L. acknowledges financial support from the Enhanced Eurotalents fellowship, a Marie Sklodowska-Curie Actions Programme.

%%%%%%%%%%%%%%%%%%
%
%

%%%%%%%%%%%%%%
\newpage
\appendix

%%%%%%%%%%%%%%%%
%
%
%
%

%%%%%%%%%%%%%%%
%
%
%
%%%%%%%%%%%%%%%
\section{IR-resummation expressions} \label{irresumexpsec}

In this appendix, we present some details for doing the IR-resummation presented in \secref{analyticresumsec}.  First, we introduce some notation.  We use a double bar $g(r)||_n$ to mean that the quantity $g(r)$ is expanded up to $n$-th order in all of the parameters $\epsilon_{s <}$, $\epsilon_{ \delta < }$, and $\epsilon_{s >}$ (defined in \eqn{epsilondef}): this is simply the Eulerian expansion.  We use a single bar $g(r)|_n$ to mean that $g(r)$ is expanded up to $n$-th order in $\epsilon_{ \delta < }$ and $\epsilon_{s >}$, but that $\epsilon_{s<}$ has been resummed: this is the result of the IR-resummation.  Finally, for the correlation function, we use a subscript like in $\xi(r)_j$ to mean that we take the $j$-th loop-order piece of the correlation function in Eulerian perturbation theory.

From \cite{Senatore:2017pbn}, the IR-resummed correlation function $\xi(r )\big|_N $, to order $N$ in $\epsilon_{\delta <}$ and to all orders in $\epsilon_{s <}$, in terms of the $j$-th order contribution to the Eulerian correlation function $ \xi^{\rm E}_j ( q )$, is given by\footnote{Here and in the rest of the paper, we only include the effects of the linear displacements; Ref. \cite{Senatore:2017pbn} showed that the correction due to the three-point function of the displacement field is negligible with respect to higher loop corrections.}
\be \label{generalresumeq}
\xi(r )\big|_N =  \sum_{j=1}^N \int d^3 q \,  \xi^{\rm E}_j ( q ) R_{N-j} ( \rvec - \qvec , \rvec)
\ee
where 
\be \label{rnj}
R_{N-j} ( y^i , \rvec)  = K_0^{-1} \left( - i \frac{\partial}{\partial y^i} , \rvec \right)\Bigg| \Bigg|_{N-j} G( \yvec , \rvec) \ ,
\ee
\be
G \left(\yvec , \rvec \right) = \frac{(2 \pi)^{-3/2}}{\sqrt{| A ( \rvec ) |}} \exp \left\{ - \half y^i  y^jA_{ij}^{-1} ( \rvec )  \right\}  \ ,
\ee
\be \label{k01}
K_0 ( \kvec , \qvec ) = \exp \left\{ - \half k^i k^j A_{ij} ( \qvec )  \right\} \ , 
\ee
\be \label{amx}
A_{ij} ( \rvec ) \equiv A_0 ( r ) \delta_{ij} + A_2 ( r ) \hat r_i \hat r_j \ , \hspace{.5in} A_{ij}^{-1} (\rvec ) \equiv \alpha_0 ( r ) \delta_{ij} + \alpha_2 ( r ) \hat r_i \hat r_j \ , 
\ee
\begin{align} \label{a0def}
A_0 ( r )  = \int_0^{\Lambda_{\rm IR}} \frac{d \, p }{2 \pi^2}   \frac{2}{3} \left( 1 - j_0 ( r p ) - j_2 ( r p ) \right) P_{11} ( p)\  , \quad \quad  A_2 ( r )  = \int_0^{\Lambda_{\rm IR}} \frac{d \, p }{2 \pi^2} 2 j_2 ( r p )  P_{11} ( p)\ , 
\end{align}
$\alpha_0 (r ) = A_0(r)^{-1}$, and $\alpha_2 (r) = - A_2(r) A_0(r)^{-1} ( A_0(r) + A_2(r))^{-1}$.   Here, $\Lambda_{\rm IR}$ is the IR scale up to which we resum the linear IR modes.  This cutoff and the choice in \eqn{k01} to keep the linear modes non-perturbative both serve to define a new expansion parameter $\tilde \epsilon_{s<} $, such that $\tilde \epsilon_{s<} \ll 1 \lesssim \epsilon_{s<}$.  Because $\tilde \epsilon_{s<}$ is small, the perturbative expansion now converges rapidly, and higher loops will improve the computation.  In practice, we use $\Lambda_{\rm IR} = 0.12 \unitsk$.

In this paper, we concentrate on the one-loop resummation, so the relevant functions for us are
\be
R_0 ( \yvec , \rvec )  = \frac{(2 \pi)^{-3/2}}{\sqrt{| A ( \rvec ) |}} \exp \left\{ - \half y^i y^j A_{ij}^{-1} ( \rvec )  \right\}  \ , 
\ee
and
\be \label{r1eq}
R_1 ( \yvec , \rvec) =  \frac{(2 \pi)^{-3/2}}{\sqrt{| A ( \rvec ) |}}  \left( \frac{5}{2} - \half y^k y^l A^{-1}_{kl} ( \rvec )  \right) \exp \left\{ - \half y^i y^j A_{ij}^{-1} ( \rvec )  \right\}   \ . 
\ee
We will see that it is useful to introduce the following notation
\be \label{rlambda0}
R^\lambda_0 ( \yvec , \rvec ) = \frac{(2 \pi)^{-3/2}}{\sqrt{| A ( \rvec ) |}} \exp \left\{ - \frac{\lambda}{2} y^i A_{ij}^{-1} ( \rvec ) y^j \right\}
\ee
so that we can write 
\be \label{r1r0exp}
R_1 ( \yvec , \rvec) = \frac{5}{2} R^1_0 ( \yvec , \rvec ) + \partial_\lambda R_0^\lambda ( \yvec , \rvec) \big|_{\lambda = 1} \ . 
\ee

Now let us briefly comment on some properties of the matrix $A_{ij} ( \rvec ) $ used in writing the integral in \eqn{tildexizero} as \eqn{quicksand}.  From the expression in \eqn{amx}, we can immediately find the three eigenvectors and eigenvalues of $A_{ij} ( \rvec)$.  Let $\hat v_i^{(1)}$ and $\hat v_i^{(2)}$ be two unit vectors that are orthogonal to $\hat r_i$.  Then $\hat v^{(1)}_i$ and $\hat v^{(2)}_i$ are eigenvectors with eigenvalues $A_0(r)$, and $\hat r_i$ is an eigenvector with eigenvalue $A_0(r) + A_2(r)$.  This means that we can diagonalize $A$ and $A^{-1}$ as 
\begin{align}
A_{ij} ( \rvec ) &\rightarrow \begin{pmatrix}  A_0(r) & 0 & 0 \\ 0 & A_0(r) & 0 \\ 0 & 0& A_0(r) + A_2 (r)  \end{pmatrix} \\
 A^{-1}_{ij} ( \rvec ) &  \rightarrow \begin{pmatrix}  \alpha_0(r) & 0 & 0 \\ 0 & \alpha_0(r) & 0 \\ 0 & 0& \alpha_0(r) + \alpha_2 (r)  \end{pmatrix} 
\end{align}
where we have chosen the $\hat 3 $ direction to be parallel with $\hat r$.

%%%%%%%%%%%%%%%%%%
%
%

%%%%%%%%%%%%%%%%%%

\section{Subleading corrections to the wiggle/no-wiggle method}  \label{wnwcxnsec}

\subsection{Derivation of wiggle/no-wiggle from the IR-resummation}
The expression for the resummed power spectrum at one loop in the wiggle/no-wiggle method is given by \cite{Baldauf:2015xfa,Vlah:2015zda}
\be \label{wnwps}
P^{\rm w/nw} ( k ) |_1 = P^s_{11} ( k ) + P^s_1 ( k ) + e^{-k^2 \Sigma^2} \left( P_{11}^w ( k ) \left( 1 + k^2 \Sigma^2 \right) + P_1^w ( k ) \right)
\ee
where $P_{11}^s(k)$ is the smoothed linear power spectrum (for example using the Eisenstein and Hu smooth power spectrum \cite{Eisenstein:1997jh}), $P_{11}^w(k) \equiv P_{11}(k) - P_{11}^s ( k ) $, $P_{1}^s(k)$ is the one-loop contribution to the power spectrum computed in Eulerian perturbation theory using only $P_{11}^s (k)$ (including counterterms), $P_1^w ( k ) \equiv P_1 ( k )  - P^s_1 ( k )$, and $\Sigma^2$ is defined differently in the two references \cite{Baldauf:2015xfa,Vlah:2015zda}.  In \cite{Baldauf:2015xfa}, calling their choice $\Sigma_1^2$, they use
\be \label{sigma1}
\Sigma_1^2 = \frac{1}{2} A_0 ( \lbao) \ , 
\ee
where $A_0$ is defined in \eqn{a0def}, but with $\Lambda_{\rm IR} = \epsilon k$ and $\epsilon \ll 1$.  On the other hand, using $\Sigma_2$ to denote the choice in \cite{Vlah:2015zda}, they use
\be \label{sigmadef}
\Sigma_2^2 = \frac{1}{2 \pi^2}  \frac{1}{r_{\rm max}^3 - r_{\rm min}^3} \int d^3 r   \left( A_0^{s,\infty} ( r ) + \frac{1}{3} A_2^{s,\infty} ( r ) \right) \ , 
\ee
which is the average of the smooth part of the displacement correlation functions over the range where the wiggles are prominent (for example $q_{\rm min} = 10 \unitsr$ and $q_{\rm max} = 300 \unitsr$), and $A_0^{s,\infty}$ and $A_2^{s,\infty}$ are as in \eqn{a0def}, but using $P_{11}^s$ and integrating with $\Lambda_{\rm IR}~=~ \infty$.  The difference between evaluating the displacements at $\lbao$ as in \eqn{sigma1} or averaging over them as in \eqn{sigmadef} is a small effect, for the same reason that the fixed-displacements approximation of \secref{fixeddispsec} is a good approximation.

 In this appendix we are going to rigorously derive \eqn{wnwps} in such a way that all of the approximations will be explicit and one could, in principle, include them.\footnote{We do this derivation at one-loop, but it is obvious how to extend it to all loops, as the trick of subtracting the IR-resummation when adding the loop conbtribution so that we are not double counting is the same as introduced in \cite{Senatore:2014via}.}  Let us first see how \eqn{wnwps} can be simply derived from the expressions of \cite{Senatore:2017pbn}, which are a simple manipulation of the expressions of \cite{Senatore:2014via}, after dropping some small corrections.  The one-loop expression for the correlation function is given by, see \appref{irresumexpsec}, 
\be \label{oneloopexpression}
\xi ( r ) |_1 = \frac{(2 \pi)^{-3/2}}{ \sqrt{|A ( \rvec ) |}} \int d^3 q \exp \left\{ - \half q^i q^j A_{ij}^{-1} ( \rvec) \right\} \left( \left( \frac{5}{2} - \half q^k q^l A_{kl}^{-1} ( \rvec) \right) \xi_{11}^{\rm E} (| \rvec - \qvec|  ) + \xi^{\rm E}_{1} ( | \rvec - \qvec | )\right) \ . 
\ee
Breaking the power spectrum into a smooth part and a wiggle part means that we break the correlation function into a smooth part and a peak part.  Similarly to above, we define $\xi_{11}^s$ and $\xi_1^s$ as the Fourier transforms of $P_{11}^s$ and $P_1^s$, and we also define $\xi^p_{11} \equiv \xi_{11}^{\rm E} - \xi_{11}^s$ and $\xi^p_{1} \equiv \xi_{1}^{\rm E} - \xi_{1}^s$, where $p$ stands for peak.  

We start by looking at the resummation \eqn{oneloopexpression} applied to the smooth correlation functions.  One can notice immediately that if $\xi^s ( | \rvec - \qvec |) \rightarrow \xi^s ( r )$ in \eqn{oneloopexpression}, which is generally a good approximation since from \cite{Senatore:2014via} we know that at leading order, the IR-resummation does not change a truly smooth function, one can do the Gaussian integral over $d^3q$ to obtain $\xi^s ( r ) |_1 \rightarrow \xi_{11}^s ( r ) + \xi_1^s ( r )$, i.e., there is no resummation of the smooth part.  We will comment below on the corrections to this statement, but first we look at the leading resummation of the peak part.  

Next, consider \eqn{oneloopexpression} applied to the peak correlation function $\xi^p$ (whose Fourier transform is the wiggle power spectrum $P^w$), and write the expression in Fourier space.  This gives\footnote{ We have used the Fourier transforms of the Gaussians
\begin{align}
\begin{split}
G( \qvec , \rvec ) & \equiv \exp \left\{ - \half q^i q^j A_{ij}^{-1} ( \rvec) \right\} = \int_{\kvec} e^{- i \kvec \cdot \qvec} \, \tilde G ( \kvec , \rvec)  \\
\tilde G ( \kvec , \rvec) & \equiv \int d^3 q e^{i\kvec \cdot \qvec} \exp \left\{ - \half q^i q^j A_{ij}^{-1} ( \rvec) \right\} = \frac{(2 \pi)^{3/2}}{ \sqrt{ | A^{-1} ( \rvec) | }} \exp \left\{ - \half k^i k^j A_{ij} ( \rvec ) \right\} \\
G_1( \qvec , \rvec ) & \equiv q^k q^l A^{-1}_{kl} ( \rvec) \exp \left\{ - \half q^i q^j A_{ij}^{-1} ( \rvec) \right\} = \int_{\kvec} e^{- i \kvec \cdot \qvec} \, \tilde G_1 ( \kvec , \rvec)  \\
\tilde G_1 ( \kvec , \rvec) & \equiv \int d^3 q \, e^{i \kvec \cdot \qvec} G_1 ( \qvec , \rvec )  = \frac{(2 \pi)^{3/2}}{ \sqrt{ | A^{-1} ( \rvec) | }} \exp \left\{ - \half k^i k^j A_{ij} ( \rvec ) \right\}  \left( 3 - k^i k^j A_{ij} (\rvec) \right)  \ .
\end{split}
\end{align}
}
\begin{align} \label{pw1exp}
P^w ( k ) |_1 &\equiv \int d^3 r \, e^{i \kvec \cdot \rvec } \xi^p(r) |_1 \\
&   = \frac{(2 \pi)^{-3/2}}{\sqrt{| A ( \rvec ) |}} \int d^3 r \, e^{i \rvec \cdot \kvec} \int_{\kvec'}  e^{-i \kvec ' \cdot \rvec} \left(\left( \frac{5}{2} \tilde G ( \kvec ' , \rvec) - \half \tilde G_1 ( \kvec ' , \rvec) \right)   P^{w}_{11} ( k')  + \tilde G ( \kvec ' , \rvec )  P^w_1 ( k')   \right) \nonumber \\
& =   \int d^3 r \, e^{i \rvec \cdot \kvec} \int_{\kvec'}  e^{-i \kvec ' \cdot \rvec} \exp \left\{ - \half k'^i k'^j A_{ij} (\rvec) \right\} \left(  \left(  1 +\half k'^i k'^j A_{ij} ( \rvec)  \right) P^w_{11} ( k' ) + P^w_{1} ( k')  \right)  \nonumber \ .
\end{align}
Now, if $A_{ij}$ were not to depend on $\rvec$, i.e. if $A_{ij} = \bar A_0  \, \delta_{ij}$ where $\bar A_0$ is constant, then one could do the integral over $d^3 r$ to get a factor of $\delta ( \kvec - \kvec')$ and then do the integral over $\kvec '$ to give
\be  \label{pwderive}
P^w ( k ) |_1 \rightarrow \exp \left\{ - \half k^2 \bar A_{0}  \right\} \left(  \left(  1 +\half k^2 \bar A_{0}   \right) P^w_{11} ( k ) + P^w_{1} ( k)  \right) \ . 
\ee
This is almost the expression for the wiggle part in \eqn{wnwps}, but we should make a few comments related to the matrix $A_{ij} ( \rvec) = A_0 ( r ) \delta_{ij} + A_2 ( r ) \hat r_i \hat r_j$.  To make this independent of $r$, one could consider a spatial average over $A_0 (r )$ and $A_2 ( r )$, as in \eqn{sigmadef}, or simply evaluate them near the BAO peak, i.e. use $A_0( \ell_{\rm BAO} )$ and $A_2 ( \ell_{\rm BAO})$ as in \eqn{sigma1} (the difference between these two choices is small).  In terms of the angular part of the matrix structure, to ignore the $\hat r_i$ dependence, one can either take the piece proportional to $\delta_{ij}$ as in \eqn{sigma1}, or take the trace, which is what \eqn{sigmadef} does.  Finally, the displacement correlation functions $A_0$ and $A_2$ that we use have a cutoff at $\Lambda_{\rm IR} = 0.12 \unitsk$, while the functions $A_0^{s,\infty}$ and $A_2^{s,\infty}$ in \eqn{sigmadef} have a cutoff of $\Lambda_{\rm IR} = \infty$, and $A_0( \lbao)$ in \eqn{sigma1} has a cutoff of $\Lambda_{\rm IR} = \epsilon k$, with $\epsilon \sim 1/2$.  With these caveats in mind, we recover the wiggle part in \eqn{wnwps} by using
\be
\bar A_0 \rightarrow A_0 ( \ell_{\rm BAO} )  \simeq 2 \Sigma_1^2  \ , 
\ee
to match \eqn{sigma1}, or 
\be
\bar A_{0} \rightarrow \left( A_0 ( \ell_{\rm BAO} ) + \frac{1}{3} A_2 ( \ell_{\rm BAO} ) \right)  \simeq 2 \Sigma_2^2  \ ,
\ee
to match \eqn{sigmadef}.  In this way have shown how the wiggle/no-wiggle method is a relatively straightforward approximation of the expression in \cite{Senatore:2014via}.  Indeed, this had to be the case as there is only one correct way to resum the displacements.  However, our derivation allows us to estimate the size of the approximations, or to include the corrections, which we do next.

%%%%%%%%%%%%%%%%%

\subsection{Subleading corrections }

\subsubsection{Smooth part}
Now that we have seen the approximations for which the IR-resummation reduces to the wiggle/no-wiggle method, we can study the corrections.  The first correction comes from the resummation of the smooth part of the correlation function.  As stated above, this comes from expanding $\xi^s ( | \rvec - \qvec |)$ for $q/r \ll 1$.  Looking back at \eqn{oneloopexpression}, plugging in the smooth correlation functions, expanding $\xi^s ( | \rvec - \qvec |)$ for $q/r \ll 1$, and looking at the first correction, we find
\be \label{deltaxis}
\Delta \xi^s ( r ) |_1 \approx  \half  \frac{ \partial^2 \xi_{1}^s}{\partial r^i \partial r^j} ( \rvec )  A_{ij} ( \rvec)  \ . 
\ee
If one has a perfectly smooth function and there is no scale in it, then one has that
\be
\frac{ \partial^2 \xi_{1}^s}{\partial r^i \partial r^j} ( r) \sim \frac{ \xi_1^s ( r ) }{r^2} \ . 
\ee
However, if during the arbitrary splitting of $\xi^{\rm E}$, there are still some peak-like features in the smooth correlation function, then
\be
\frac{ \partial^2 \xi_{1}^s}{\partial r^i \partial r^j} ( r ) \sim \frac{ \xi_1^s ( r ) }{r^2} + \epsilon_{\rm osc} \frac{ \xi_1^s ( r ) }{\sigma^2} \ , 
\ee
where $\sigma$ is the size of the feature near position $q$ in configuration space (for example the width of the BAO peak near $\lbao$), and $\epsilon_{\rm osc}$ characterizes the oscillatory features that are still present in the power spectrum.  We will discuss these two corrections separately, so for convenience, let us define
\be \label{correctiondefs}
\Delta \xi^s_{1, \text{smooth}} (r) \equiv \half | A  (r)| \frac{\xi^s_1 ( r ) }{r^2 } \ , \hspace{.1in} \text{and,} \hspace{.1in} \Delta \xi^s_{1, \text{osc}} (r) \equiv \half | A  (r)| \epsilon_{\rm osc} \frac{ \xi^s_1 ( r ) } {\sigma^2 }  \  ,
\ee
where by $| A(r) |$ we mean the typical size of the matrix $A_{ij} ( \rvec )$, which for example, we can take to be $| A ( r ) | = A_0 ( \lbao) $ for the purposes of this discussion.

\begin{figure}[htb!]
\includegraphics[width=8cm]{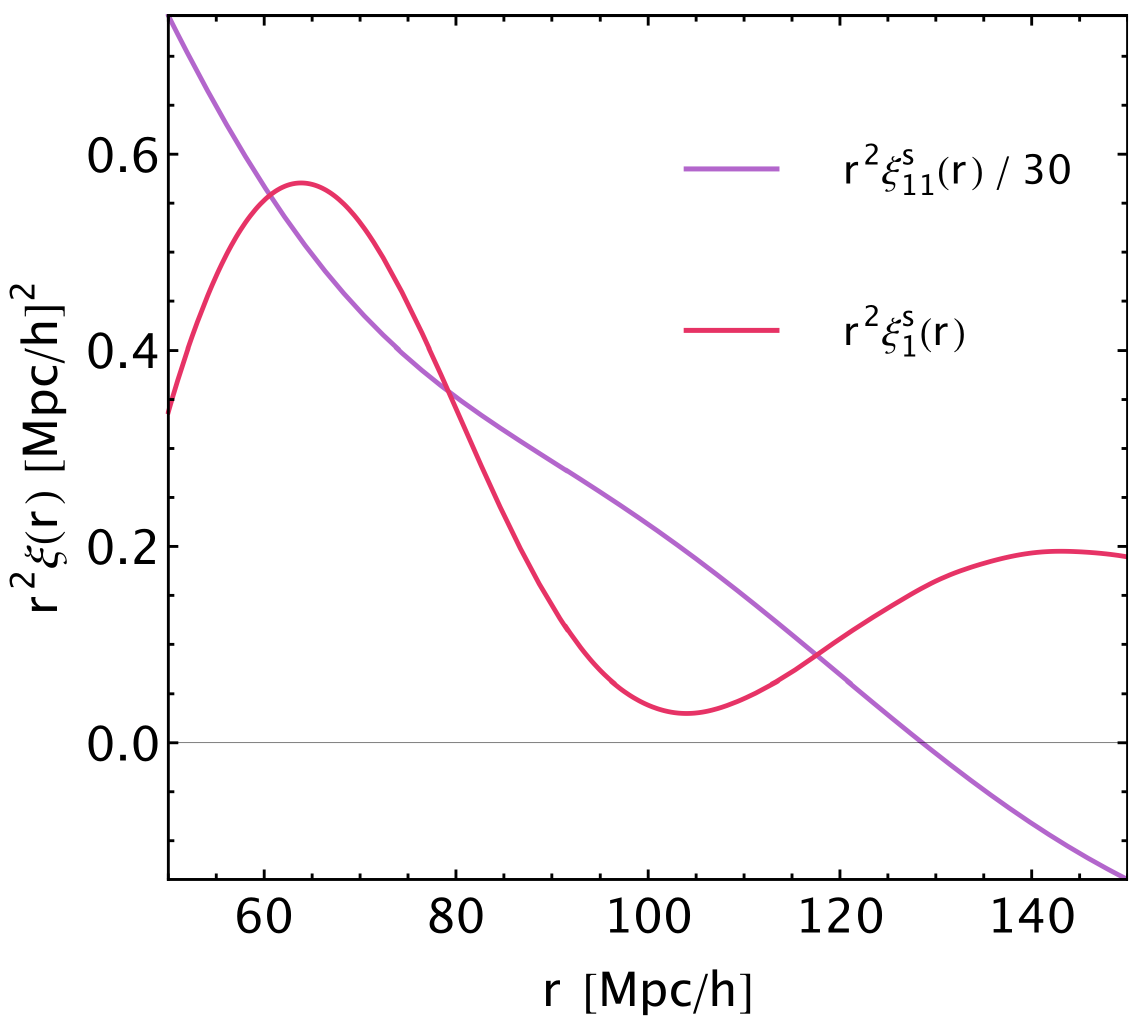} \includegraphics[width=8.5cm]{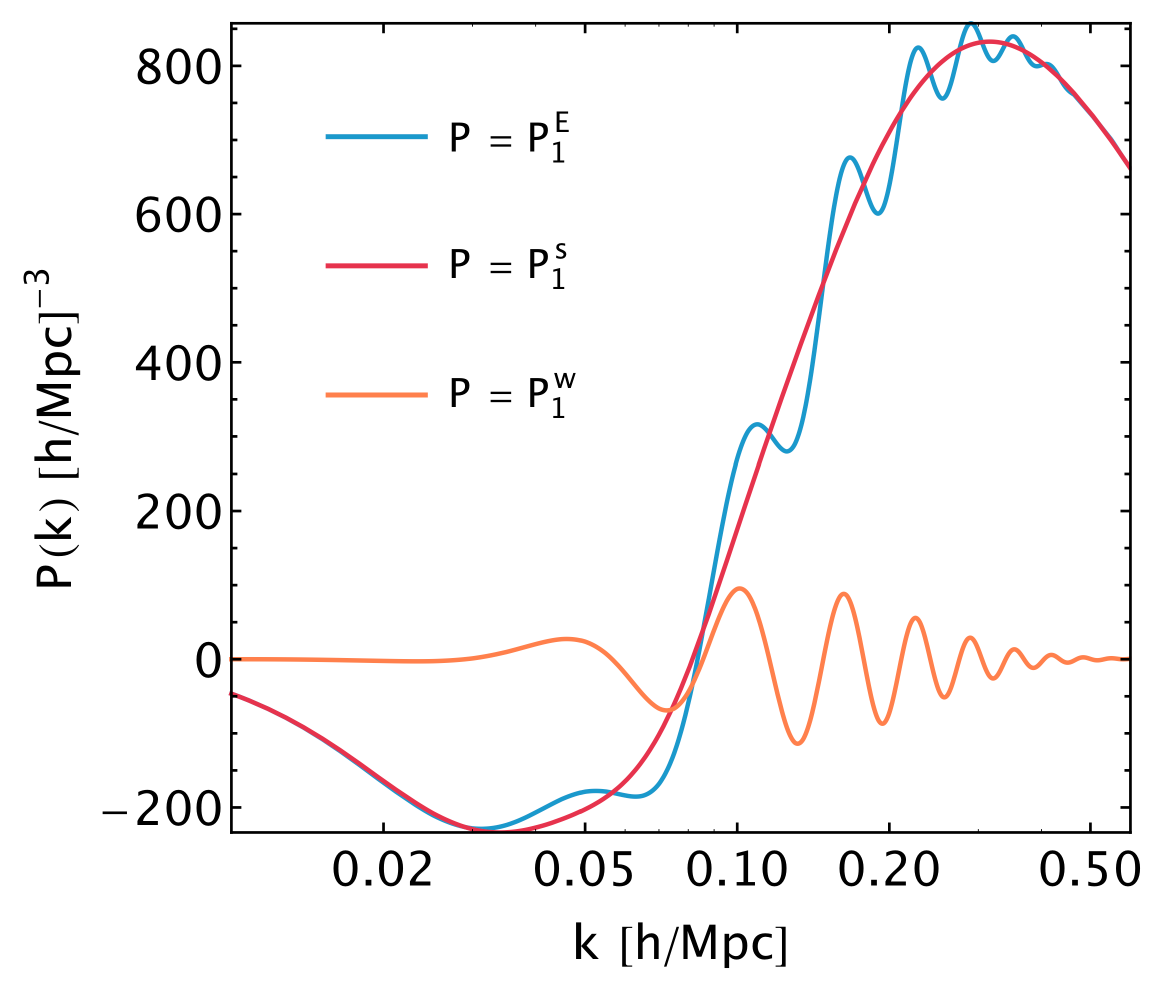}  \caption{In this figure, in the left panel, we show the smoothness of the tree level $\xi^s_{11}$ and one-loop contribution $\xi^s_{1}$, and we see that, while $\xi_{11}^s$ is quite smooth as expected, $\xi_1^s$ is much less so.  The reason that $\xi_1^s$ is not smooth is that there is a feature in the one-loop contribution to the power spectrum, even in the absence of baryons, for $k < 0.5 \unitsk$, as can be seen in the plot of $P^s_1$ in the right panel (red curve).   } \label{xinotsmooth}
\end{figure}

We first discuss the correction $\Delta \xi^s_{1, \text{smooth}} $.  It turns out that the size of the displacements $A_{ij}$ at the BAO scale are of order the BAO peak width squared, $\sigma^2$, so $\Delta \xi^s_{1, \text{smooth}} ( \lbao)$ is a parametrically small correction if $\sigma^2 / \lbao^2 \ll 1$, which is more or less true since $\sigma \approx 10 \unitsr$ and $\lbao \approx 110 \unitsr$.\footnote{The error is also small away from the BAO peak, i.e. for $r \ll \lbao$, because $| A( r ) | / r^2 \simeq 0.014$ as $r \rightarrow 0$.}  This may not always be the case with $\Delta \xi^s_{1, \text{osc}}$, though.  Looking at \figref{xinotsmooth}, we see that $\xi^s_1$ has features somewhat comparable to the scale of the BAO-peak width.\footnote{A less optimal choice of $\xi^s$ might put into $\xi^s$ also some of the oscillatory features of the BAO. In fact, clearly, if one were to choose, very suboptimally, $\xi^s \equiv \xi^{\rm E}$, then the wiggle/no-wiggle procedure would be equivalent to SPT.}  These features come from the fact that the one-loop contribution to the power spectrum, even in the absence of baryons, has an oscillatory-like feature before $k \approx 0 .5 \unitsk$.  As can be seen in \figref{xicxns1}, this leads to an approximately $0.5 \%$ correction near the BAO peak.  In \figref{xicxns1}, we also see that $\Delta \xi^s_{1, \text{smooth}} $ is indeed negligible, and that the extra features in $\xi^s_1$ make $\epsilon_{\rm osc} \approx 2.5$.

\subsubsection{Peak part}

We now turn to the corrections to the resummation of the peak correlation functions, which turn out to be comparable to the corrections found in the previous section.  Recall that to get the wiggle/no-wiggle expression in \eqn{pwderive}, we had to assume that $A_{ij}$ was independent of $\rvec$.  The corrections to that assumption come from two sources: the first is the angular dependence in the $\hat r_i \hat r_j$ term, and the second is the magnitude dependence in $A_0(r)$ and $A_2(r)$.  It turns out that the angular dependence is most important, so we concentrate on that here.

\begin{figure}[htb!]
\hspace{.4in} \includegraphics[width=14cm]{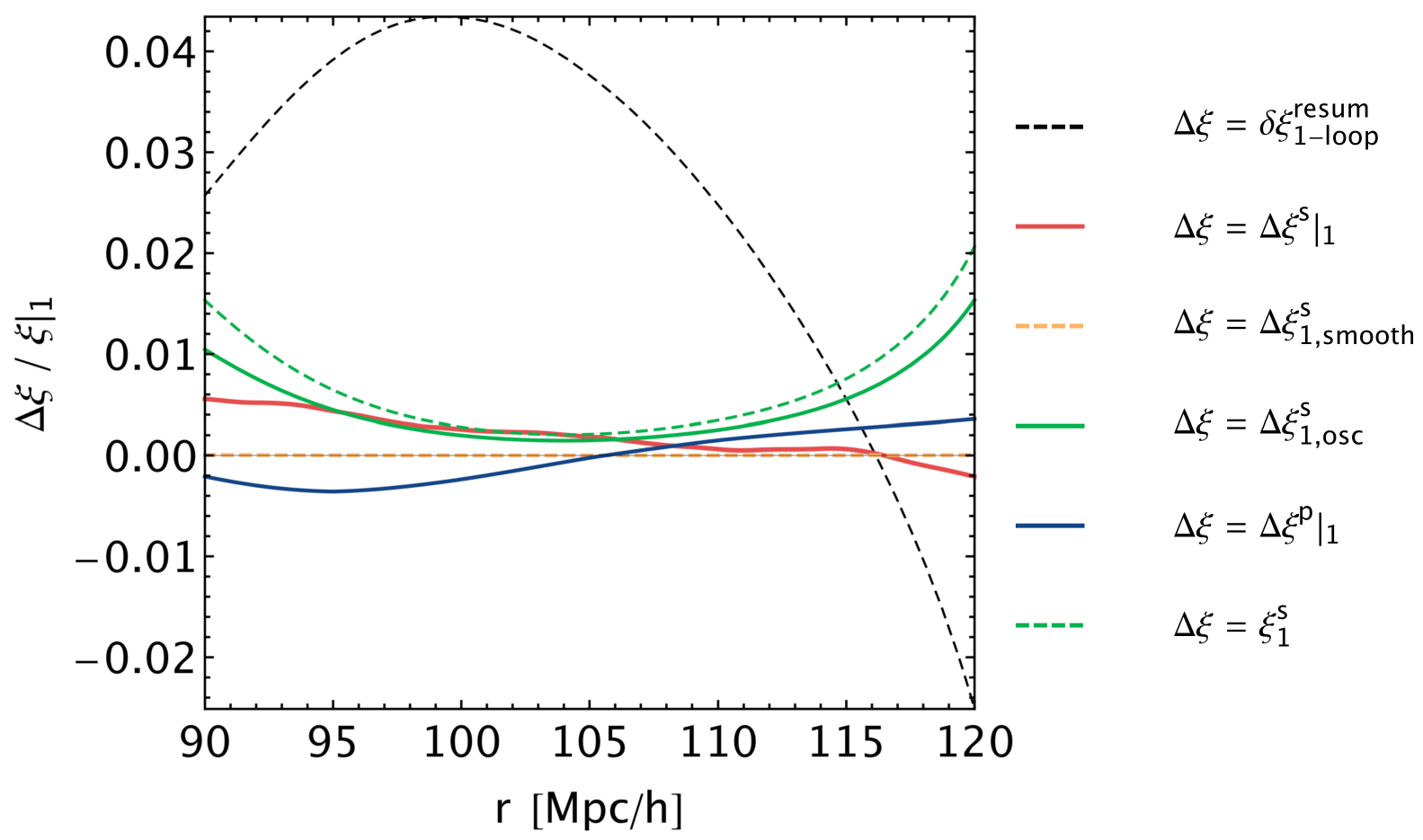}  \caption{In this figure, we plot various corrections to the wiggle/no-wiggle method.  In the legend, $\delta \xi_{1\text{-loop}}^{\rm resum}$ is the change in the IR-resummation going from tree-level to one-loop, $\Delta \xi^s |_1$ is defined in \eqn{deltaxis}, $\Delta \xi_{1,\text{smooth}}^s$ and $\Delta \xi_{1,\text{osc}}^s$ are defined in \eqn{correctiondefs}, $\Delta \xi^p |_1$ is defined in \eqn{peakcxn}.  In this figure, we use $\epsilon_{\rm osc} = 2.5$ to show that the correction to the resummed smooth one-loop contribution comes from the fact that $\xi_1^s$ is not totally smooth and has sizeable features, even though the estimate of the correction from the smooth correlation function (the dashed orange curve, which is barely different from $0$ in this plot), is negligible.  We have also included the smooth one-loop contribution $\xi^s_1$ as the dashed green curve for comparison.  It is however possible, and somewhat suggested by the plot, that these independent corrections (the red and blue curves) might accidentally cancel, at least partially.  } \label{xicxns1}
\end{figure}

To find the corrections due to the angular dependence, we write
\be
A_{ij} ( \rvec ) \approx \left( A_0 ( \lbao ) + \frac{1}{3} A_2 ( \lbao) \right) \delta_{ij} + A_2 ( \lbao) \left( \hat r_i \hat r_j - \frac{1}{3} \delta_{ij} \right) + \dots \ ,
\ee
and expand $A_{ij}$ as above in the last line of \eqn{pw1exp}, keeping only the terms which are first order in $A_2 ( \lbao) ( \hat r_i \hat r_j - \delta_{ij}/3)$.  The correction due to the angular terms then, is
\begin{align} \label{peakcxn}
\begin{split}
\Delta \xi^p ( r ) |_1 & = \int_{\kvec} e^{- i \kvec \cdot \rvec} \exp \left\{ - \half k^2   \left( A_0(\ell_{\rm BAO} ) + \frac{1}{3} A_2( \ell_{\rm BAO})  \right) \right\}  \left( - \frac{k^2 A_2(\ell_{\rm BAO})}{2} \left( \mu^2 - \frac{1}{3} \right) \right) \\ 
& \quad \quad \quad \times \left( \half k^2   \left( A_0(\ell_{\rm BAO} ) + \frac{1}{3} A_2( \ell_{\rm BAO})  \right) P^w_{11} ( k ) + P^w_1 ( k )  \right) \ ,
\end{split}
\end{align}
where $\mu \equiv \hat k \cdot \hat r$.  To see how large of a correction this might be, we should compare it to the leading term in \eqn{pw1exp}, which is 
\be
\approx  \int_{\kvec} e^{- i \kvec \cdot \rvec} \exp \left\{ - \half k^2   \left( A_0(\ell_{\rm BAO} ) + \frac{1}{3} A_2( \ell_{\rm BAO})  \right) \right\} P_{11}^w ( k) \ . 
\ee
To estimate the difference in the integrals over $\mu$, we use that 
\be
\frac{ \int_0^\infty d x \int_{-1}^1 d\mu \, e^{-i x \mu}}{   \int_0^\infty d x \int_{-1}^1 d\mu \, e^{-i x \mu} \left( \mu^2 - \frac{1}{3}\right) } \approx -3 \ , 
\ee
so that the size of the correction is approximately
\be \label{correction1}
\left| \frac{\Delta \xi^p ( r ) |_1}{\xi^p ( r ) |_1} \right| \simeq \frac{\frac{1}{4} A_2 ( \lbao) k^4 \left( A_0 ( \lbao) + \frac{1}{3} A_2 ( \lbao) \right) }{ 3} \approx 0.009
\ee
where we have used $k = 0.1 \unitsk$, which is where the BAO wiggles have support.  We see in \figref{xicxns1} that our estimate is of the correct order.

We should note that, at higher loop order, the effect of some of these corrections is expected to diminish because of the analytic dependence of the power spectrum on $\esl$ but that different corrections will scale differently.  For example, differences related to the cutoff $\Lambda_{\rm IR}$ or to the difference between \eqn{sigma1} and \eqn{sigmadef} with respect to evaluating the displacements at $\lbao$ or averaging over them, will be parametrized by $\tilde{\epsilon}_{s <} \ll \esl \simeq 1$ and are recovered in the SPT expansion (this is analogous to the discussion in \secref{fixeddispsec} about the fixed-displacements approximation).  On the other hand, differences related to the angular expansion are parametrized by the smallness of $A_2 ( \lbao ) / \lbao^2$ times small angular factors, which happen to be small in practice, but are generically of the order $\esl$.  However, it is not clear how the corrections due to not resumming the higher order smooth parts scale, since this depends on how one defines the smooth parts.

In summary, we have derived the wiggle/no-wiggle formula from the IR-resummation expressions of \cite{Senatore:2014via}, including the two numerically leading corrections to the wiggle/no-wiggle method, due to not resumming the smooth correlation function and to ignoring angular pieces, which are each roughly $0.5\%$ in size.  However, it is possible that there is a partial accidental cancellation among these corrections.  These corrections should be compared with the size of the two-loop correction.  By using the method proposed in \cite{Senatore:2014via} to optimize convergence (i.e. rescaling $A_0 \rightarrow ( 1 + \alpha ) A_0$ for $\alpha \sim 0.85$), the two-loop correlation function is about $0.4\%$ of one loop, which appears to be smaller, although not by much, than the estimated corrections to the wiggle/no-wiggle method.  Given that the error bars on upcoming surveys will be at the $1\%$ level~(see for example \cite{2012MSAIS..19..314C}), it is interesting to understand the differences more thoroughly.  Also, since one of the errors is related to the angular pieces, it could be that the difference will be more severe in other observables like the bispectrum and in redshift space, where these terms are expected to contribute at leading order to higher moments.  In order to do a useful and thorough comparison, we would need precise data for the BAO peak, which is not currently available to us.  However, since the main goal of this paper is to develop an analytic implementation of the formulas of \cite{Senatore:2014via}, we leave this comparison for future work.  

%%%%%%%%%%%%%%%
%
%
%
%%%%%%%%%%%
\section{One-loop power spectrum} \label{oneloopformsec}

In this appendix, we give some general background formulas that are used in this paper.  In the EdS approximation, the power spectrum up to one loop is given by (for example, see \cite{Bernardeau:2001qr})
\be \label{pexpand}
P(k,a) = D(a)^2 P_{11} ( k ) + D(a)^4 \left( P_{13} ( k ) + P_{22} ( k) \right) - 4 \pi c_s^2 ( a ) k^2 D(a)^2 P_{11} ( k )
\ee
where the last term is the counterterm from the EFTofLSS \cite{Baumann:2010tm, Carrasco:2012cv}.  The explicit forms of the one-loop pieces are 
\begin{align}
\begin{split} \label{loopexpressions}
P_{22} ( k ) & = 2 \int \momspmeas{q} F_2 ( \qvec , \kvec - \qvec)^2 P_{11} ( q ) P_{11}( | \kvec - \qvec|) \ , \\
P_{13}(k) & = 6 P_{11} ( k ) \int \momspmeas{q} F_3 ( \qvec , - \qvec , \kvec) P_{11}(q) \ ,
\end{split}
\end{align}
where the $F_n$ are known kernels \cite{Bernardeau:2001qr}, and, in particular,
\be
 F_2 ( \qvec , \kvec - \qvec ) = \frac{5}{14} + \frac{3 k^2}{28 q^2} + \frac{3 k^2}{28 | \kvec - \qvec |^2}  - \frac{5 q^2}{28 |\kvec - \qvec|^2} - \frac{5 |\kvec - \qvec|^2}{28 q^2} + \frac{k^4}{14 |\kvec - \qvec |^2 q^2}    \ .
\ee
and
\begin{align}
\begin{split}
F_3 ( \qvec , - \qvec , \kvec ) = & -\frac{97}{1512} + \frac{\diffsq}{24 k^2} + \frac{1195 k^2}{6552 \diffsq} - \frac{19 |\kvec - \qvec|^4 }{504 q^4} + \frac{ \diffsq k^2 }{14 q^4} - \frac{5 k^4}{168 q^4} \\
& - \frac{k^6}{252 \diffsq q^4}  + \frac{211 \diffsq}{1512 q^2} - \frac{| \kvec - \qvec|^4}{72 k^2 q^2} - \frac{187 k^2 }{1512 q^2} -\frac{k^4}{504 \diffsq q^2} \\
&  - \frac{19 q^2}{504 \diffsq} - \frac{q^2}{24 k^2} + \frac{q^4}{72 \diffsq k^2} \ ,
\end{split}
\end{align}
where in the expression for $F_3 ( \qvec  , - \qvec , \kvec)$ we have changed variables of integration from $\qvec \rightarrow - \qvec$ in the terms which normally contain $| \kvec + \qvec|$ so that all terms are in the same form.

Next, let us comment briefly on some properties of the dimensional regularization expression \eqn{dimreg} and subtleties mentioned in \cite{Simonovic:2017mhp} related to using \eqn{dimreg} for the loop integrals.  The first thing to notice is that for any purely power law divergence, \eqn{dimreg} gives zero, that is, $\textsf{I} ( \nuo , \nut) = 0$ if $\nuo=0$ or $\nut=0$.  This is typical when one uses dimensional regularization.  On the other hand, if an integral has a divergence, but also has a finite part, then \eqn{dimreg} gives the finite part.  For example, consider $\nuo = \nut = 1/2$.  In this case, the UV limit of the integrand is $d^3 q /q^2$ and the integral is linearly divergent in the UV, but \eqn{dimreg} gives a finite answer because $\textsf{I} ( 1/2 , 1/2) = - 1/ (4 \pi^2)$.  To understand this, consider doing the following part of the loop calculation with a UV cutoff $\Lambda$,
\be \label{finitepiece}
\int^\Lambda \momspmeas{q}  \frac{1}{q |\kvec - \qvec|} = \int^\Lambda \momspmeas{q} \left(  \frac{1}{q |\kvec - \qvec|} - \frac{1}{q^2} \right) + \int^\Lambda \momspmeas{q} \frac{1}{q^2} \ .
\ee
In this expression, the integral over the parentheses on the right-hand side is manifestly convergent in the UV because we have explicitly subtracted out the UV divergent piece.  Then, if $\Lambda$ is larger than all of the other mass scales, we can use the dimensional regularization expression \eqn{dimreg} to approximate the integral over the parentheses on the right hand side by writing
\be \label{explaindimreg}
\int^\Lambda \momspmeas{q} \left(  \frac{1}{q |\kvec - \qvec|} - \frac{1}{q^2} \right) = \int^\infty \momspmeas{q} \left(  \frac{1}{q |\kvec - \qvec|} - \frac{1}{q^2} \right) - \int_{\Lambda}^\infty \momspmeas{q} \left(  \frac{1}{q |\kvec - \qvec|} - \frac{1}{q^2} \right) \ 
\ee
and using \eqn{dimreg} to evaluate the $\int^\infty$ term above, which is the finite piece that we mentioned before, the one proportional to $\textsf{I}(1/2,1/2)$.  Because the integrand in \eqn{explaindimreg} is convergent in the UV, the piece $\int_\Lambda^\infty$ in \eqn{explaindimreg} goes to zero as $\Lambda \rightarrow \infty$, and so dimensional regularization \eqn{dimreg} gives a very good approximation to the $\int^\Lambda$ piece which we wanted to evaluate and use in \eqn{finitepiece}. Now it is clear that if we wished to use dimensional regularization to match the numerical value of the loop integral with a cutoff, we must add the last term in \eqn{finitepiece} by hand, explicitly putting the cutoff  (and, for very accurate precision, also the last integral in \eqn{explaindimreg}, which however is very small for $\Lambda \rightarrow \infty$).  This is a general lesson: if we try to do an integral which has an IR or UV divergence, \eqn{dimreg} will not pick up the divergent piece, and we should add it by hand as on the right-hand side of \eqn{finitepiece}.  We will see how this enters the loop computations a bit more concretely next.

 Now we discuss the convergence properties of the loop integrals \eqn{loopexpressions}.  When using the decomposition \eqn{plindef}, the convergence of the loop integrals depends on $\nu$, the real part of $- 2 \nu_m$ in \eqn{plindef}, so for the purposes of this discussion, let us assume that the linear power spectrum is a single power law, i.e. $P_{11} ( k ) \propto k^\nu$ where $\nu$ is real (properties of IR and UV divergences in power-law cosmologies have been thoroughly discussed in the literature, see for example \cite{Jain:1995kx, Scoccimarro:1995if, Pajer:2013jj, Carrasco:2013sva}.)  By taking the IR limit ($q \ll k$) and UV limit ($q\gg k$) of \eqn{loopexpressions} with $P_{11} ( k ) \propto k^\nu$, one can see that the $P_{22}$ integral is convergent in the IR for $-1 < \nu$ and convergent in the UV for $\nu< 1/2$, and the $P_{13}$ integral is convergent in the IR for $-1 < \nu$ and convergent in the UV for $\nu > -1$ (i.e. it is never convergent).  For concreteness, let us momentarily focus on the $P_{22}$ integral.  If we choose a value of $\nu$ such that the full integral is convergent, then using \eqn{dimreg} will certainly give us the same answer as the numerical integration.  Notice that if we expand $F_2 ( \qvec , \kvec - \qvec)^2$ and write the $P_{22}$ integral as a sum of terms of the form \eqn{dimreg}, many of the individual terms will have IR or UV divergences.  However, as mentioned before, these divergences are set to zero by dimensional regularization, and one is left with only the finite contributions which indeed add up to the final finite answer for the convergent integral.  If, on the other hand, we choose a value of $\nu$, say $\nu=1$, for which the full $P_{22}$ integral really contains a divergence (in this case there is a linear divergence in the UV), then this piece will not be captured by dimensional regularization, and as discussed around \eqn{finitepiece}, one has to add this piece by hand.  In equations, this is
\be
2 \int^\Lambda \momspmeas{q} F_2 ( \qvec , \kvec - \qvec)^2 q^\nu  | \kvec - \qvec|^\nu \simeq k^3  k^{2\nu} M_{22} ( -1/2,-1/2 ) + 2 \int^\Lambda \momspmeas{q} \frac{9}{196} \frac{k^4}{q^4} q^{2 \nu} \ , 
\ee
 where the $\simeq$ is used above because the left-hand side is equal to the right-hand side up to the integral from $\Lambda$ to $\infty$ of a UV convergent integrand, as discussed near \eqn{explaindimreg}.

Let us now discuss how this all plays out for the one-loop power spectrum in the EFTofLSS.  We start with the UV divergences.  As mentioned above, $P_{13}$ is UV divergent for $\nu > -1$.  For $-1 < \nu < 1$, there is only one term that is set to zero by dimensional regularization and that we have to add by hand.  This term is 
\be \label{p13uv}
-\frac{61}{630 \pi^2} k^2 P_{11} ( k ) \int_0^\Lambda dq \, P_{11} ( q ) \ ,
\ee
where $\Lambda$ is the UV cutoff in the EFTofLSS.  Notice that this term is of exactly the same form, as a function of $k$, as the counterterm which is already included in \eqn{pexpand}.  This is not an accident, since the counterterms in the EFTofLSS are designed to absorb the cutoff dependence of the loop integrals in the UV.  Thus, if we include the counterterm from the beginning, which has the free coefficient $c_s^2 (a)$, there is no need to add the term in \eqn{p13uv}, since this simply adjusts an already unknown coefficient.  Indeed, this logic holds for \emph{all} UV divergent pieces, and at higher loops, since they will \emph{already} be included in the EFTofLSS expansion.\footnote{There is also a leading UV divergence in the $P_{22}$ diagram.  For $1/2 < \nu < 3/2$, the only UV divergent piece to be added to $P_{22}$ is 
\be \label{p22uv}
\frac{9}{196 \pi^2} k^4 \int_0^\Lambda dq \frac{P_{11} ( q ) }{q^2} \ .
\ee
This term is related to the stochastic counterterm in the EFTofLSS, and is typically subleading in a one-loop computation, which is why it is not included in \eqn{pexpand}.  In any case, following the above discussion, as long as one includes the $k^4$ counterterm consistently in the EFT expansion, one does not have to explicitly add \eqn{p22uv} to the dimensional regularization computation. 
}

Finally, we move on to the IR properties of the full $P_{13} + P_{22}$.  As discussed above, $P_{13}$ and $P_{22}$ are separately convergent in the IR for $-1 < \nu$.  However, due to the equivalence principle, the sum $P_{13} + P_{22}$ is actually convergent in the IR for $-3 < \nu$ \cite{Jain:1995kx,Scoccimarro:1995if,Peloso:2013zw,Carrasco:2013sva}.  Thus, if one uses a linear power spectrum with $-3 < \nu < -1$, there is no need to add the IR divergent pieces separately to $P_{13}$ and $P_{22}$ because they will cancel in the full sum anyway \cite{Simonovic:2017mhp}.  However, if one uses $\nu < -3$, there are subleading IR divergences which are not guaranteed to cancel in $P_{13} + P_{22}$, and so one would have to add these pieces by hand, since dimensional regularization also sets IR divergences to zero.

%%%%%%%%%%%%%
%
%
%
%%%%%%%%%%%%%
\section{Higher loops} \label{higherloopssec}

The generalization of the analytic IR-resummation to higher loop expansions of the power spectrum is straightforward once the decomposition of the power spectrum, analogous to \eqn{p1loopdecomp}, is known (relevant formulas for the two-loop power spectrum have been given in \cite{Simonovic:2017mhp}).  It is clear that a term in the expression for the $N$-loop power spectrum, which contains $N+1$ factors of the initial power spectrum $P_{11}$, will have $N+1$ general complex parameters $\nu_{m}$.  Each individual term will also have additional integer parameters related to the kernel of the specific diagram.  For example, the basic two loop integral, analogous to \eqn{dimreg}, is \cite{Simonovic:2017mhp}
\be \label{twoloopdecomp}
\int \momspmeas{q} \frac{1}{q^{2 \nu_{m_4}} | \kvec - \qvec |^{2 \nu_{m_5}}} \int \momspmeas{p} \frac{1}{ p^{2 \nu_{m_1}} |\kvec - \pvec |^{2 \nu_{m_2}} | \qvec - \pvec |^{2 \nu_{m_3}} }  \ ,
\ee
but only three of the $\nu_{m_i}$ are general complex numbers.  For concreteness, we call $I$ the total number of exponents that are needed to describe a given diagram, i.e. $I=5$ for the expression in \eqn{twoloopdecomp}.  Then, a general $N$-loop Eulerian term will have the form
\be \label{generalnloop}
P^{\rm E}_{N} ( k ) \sim \sum_{ \{m_i\}} c_{m_1} \cdots c_{m_{N+1}}  k^{-2 \nu_{m_1\cdots m_I}} M_N( \nu_{m_1} , \dots , \nu_{m_I} )
\ee
where the sum is over each of the indices $m_1 , \dots , m_I$, the matrix $M_N$ gives the information of the specific kernel in the diagram, and $\nu_{m_1\cdots m_I} \equiv \nu_{m_1} + \dots + \nu_{m_I}$.  Here, and in the rest of this section, we will not keep track of all of the factors in each expression: we simply wish to show the general form of the result, and leave a more careful presentation for future work.  Although numerical implementation of higher-loop expressions analogous to \eqn{generalnloop} may be challenging using the FFTLog decomposition, this will be the general form.  We leave a thorough study of how feasible this approach is for higher loops to future study.

Next, in order to use the resummation formula \eqn{generalresumeq}, we must convert the power spectrum to the correlation function.  Now that we have the general form \eqn{generalnloop}, we see that the correlation function takes an equally simple form
\be \label{generalcorrfn}
\xi_{N}^{\rm E} ( r )  \sim \sum_{ \{m_i\}} c_{m_1} \cdots c_{m_{N+1}} r^{2 \nu_{m_1 \cdots m_I}} \tilde M_N( \nu_{m_1}, \dots , \nu_{m_I}) \ , 
\ee
where $\tilde M_N$ is related to $M_N$ through some combination of $\Gamma$ functions coming from the Bessel transform in \eqn{besseltransf}.  

Now, we apply the formulas of \secref{irresumexpsec} to \eqn{generalcorrfn}.  Again, we will not keep track of all terms and coefficients, just the general form of the new terms.  Since, as a function of $r$, all of the $N$-loop correlation functions \eqn{generalcorrfn} are the same (a sum over complex power laws), there is nothing new about the form of the general $N$-loop correlation function.  Because of this, we will focus on the new kernel $R_{J} ( \rvec - \qvec , \rvec )$ (where $J$ runs over the relevant values of $N-j$) applied to a general complex power $q^{2 \omega}$
\be
\int d^3 q \, R_{J} ( \rvec - \qvec , \rvec )  \, q^{2 \omega} \ . 
\ee
Using \eqn{rnj} - \eqn{k01}, we see that contained in this term is 
\be
 \tilde \xi^{(J)}_\omega ( r ) \equiv \partial_\lambda^{J} \int d^3 q \, R_0^\lambda ( \rvec - \qvec , \rvec ) \, q^{2 \omega} = \partial_\lambda^{J}  \tilde \xi^\lambda_\omega ( r ) \big|_{\lambda = 1}   \ ,
\ee
where $\partial^{J}_\lambda \equiv \partial^J / \partial \lambda^J$, $R^\lambda_0$ is given in \eqn{rlambda0}, and $\tilde \xi^\lambda_\omega$ is given in \eqn{xitildelambda}.  The $q_1$ and $q_2$ integrals in the definition \eqn{xitildelambda} for $\tilde \xi^\lambda_\omega$ can be done for general $\lambda$, and,  after expanding the $\Gamma$ function as in \eqn{gammaexp1}, one is left with
\be
\tilde \xi^\lambda_\omega ( r ) = \sqrt{ \frac{\alpha_0 + \alpha_2 }{2 \pi} } \sum_j \gamma_j ( \omega) \left( \frac{2 }{\alpha_0} \right)^j   \int_{-\infty}^{\infty} d q_3 \, ( r - q_3)^{2 ( \omega - j)} \lambda^{-j-1} \exp \left\{ - \frac{\lambda}{2} ( \alpha_0 + \alpha_2 ) q_3^2 \right\} \ . 
\ee    
Now, we have to take $J$ number of derivatives with respect to $\lambda$.  There will be many terms, with coefficients that depend on $j$ and are easily calculable using the above expression, but we simply write the general term and ignore the coefficients here.  The form of the resulting integral depends on how many times the derivative hits the exponent, which we call $J'$, and the integral to be done is
\be
\int_{-\infty}^{\infty} d q_3 \, ( r - q_3)^{2(\omega - j) } q_3^{2 J'} \exp \left\{ - \frac{1}{2} ( \alpha_0 + \alpha_2 ) q_3^2 \right\} \ .
\ee
The final step is to write $q_3^{2 J'} = \left( r- ( r - q_3 )  \right)^{2 J'}$, expand this in powers of $r - q_3$ and $r$, and write the result in terms of the basic function $I_\omega ( r )$ which is defined in \eqn{iomegaint} and whose exact expression is given in \eqn{exacti}.  In this way, we see that the general IR-resummation of an $N$-loop power spectrum can be written in terms of a single function $I_\omega ( r )$.  The main challenge, of course, is that this function must be evaluated for many values of $\omega$, approximately $\mathcal{O}\left( N_{\rm max}^{N+1} \right)$, i.e. for all of the terms in the sum in \eqn{generalcorrfn}.   We have discussed in the main text how to do this for the one-loop result.

%%%%%%%%%%%%%%%
%
%

\section{Recursion relations} \label{rrapp}

In this appendix, we would like to mention a general simplifying relation which can be used in any of the strategies presented in this paper.  As can be seen from \eqn{xitildengamma}, \eqn{xi1eq}, and \eqn{jexpression}, the non-trivial functions that we have to evaluate are $I_{\omega_m +n/2}$, where $n$ is an integer.  For the one-loop computation presented above, one needs to evaluate for $n = \{-1,-1/2, 0 , 1/2, 1 \} $, but if one wishes to go to higher order in the $\Gamma$ function expansion, or to higher loops, then one will have to evaluate for more $n$.  Naively, this means that for each $r$ and $n$, we have to evaluate $I_{\omega_m +n/2} (r)$ for $N_{\rm max}$ number of $\eta_m$.  However, we can reduce the number of evaluations by using recursion relations that are satisfied by the hypergeometric function $U ( a , b ;z)$ that enters the definition of $I_{\omega_m }$ in \eqn{iomegaint}.  In particular, one recursion relation connects functions with an integer difference in the first argument $a$:
\be \label{recurrence1}
U( a  , b ; z) + ( b - 2 a - 2 -z) U(a+1,b;z) + (a+1) ( a-b+2) U( a+2 , b ;z ) = 0 \ ,
\ee
where ultimately we will set $a = -\omega$, $b = 1/2$ and $z = - r^2 ( 1 + i \epsilon)^2 ( \alpha_0 ( r ) + \alpha_2 ( r ) ) / 2$.  Because this relation involves three terms, it means that we have to initially evaluate two of them, and then the recursion relation determines all of the rest.  We have to do this once for the integer shifts in $\omega_m$, and also once for the half-integers shifts in $\omega_m$, so it means we have to evaluate a total of four shifts.  For example, we can evaluate $I_{\omega_m + n/2} ( r ) $ for $n = 0,1/2,1,3/2$ and for all of the $m$ and $r$ of interest, and then use the relation \eqn{recurrence1} to determine $I_{\omega_m +n/2} ( r ) $ for all of the other relevant $n$.  While this is not so important for the order at which we work, as it reduces the number of computations by just $\sim 20\%$, it means that it is essentially trivial to go to higher orders in the expansion of the $\Gamma$ function that enters \eqn{xitildengamma}, and also, as we will discuss, to go to higher loops, as no evaluations of the $U$ functions are needed beyond the ones that we do at one loop.

To solve the recursion relation, one can write it in matrix form:
\be
 \begin{pmatrix} U(a+2 , b ; z)  \\ U( a+1 , b ; z)   \end{pmatrix}  = K [a] \begin{pmatrix} U(a+1 , b ; z)  \\ U( a , b ; z)  \end{pmatrix}  \ ,
\ee
where $K[a]$ is a $2\times2$ matrix with elements $K[a]_{11} = -( b - 2 a - 2 - z )/ [(a+1)(a -b +2 )]$, $K[a]_{12} = -1/ [(a+1)(a -b +2 )]$, $K[a]_{21} = 1$, and $K[a]_{22} = 0$.  Then, one can easily solve for the $k$-th term by writing
\be
\begin{pmatrix} U(a+k , b ; z)  \\ U( a+k-1 , b ; z)   \end{pmatrix} = K[a+k-2] \,  K[a+k-3] \cdots K[a] \begin{pmatrix} U(a+1 , b ; z)  \\ U( a , b ; z)  \end{pmatrix} \  . 
\ee

\newpage

 \bibliographystyle{utphys}
 \small
\bibliography{matt_master_bib}

\end{document}